\documentclass[preprint,12pt]{elsarticle}
\usepackage{color}
\usepackage{graphics}
\usepackage{amssymb}
\journal{Physics Reports}

\def\ad{amplitude death}

\def\le{lyapunov exponent}
\def\ep{\epsilon}

\def\ie{i.e.~}
\def\eg{e. g.~}
\def\etl{$et ~al.$~}

\def\w{{\omega}}
\def\epsilon{\varepsilon}

\def\beq{\begin{equation}}
\def\beqr{\begin{eqnarray}}
\def\eqnr{{\end{eqnarray}\noindent}}
\def\eqn{{\end{equation}\noindent}}

\def\pre#1#2#3{{ Phys. Rev. E} {\bf #1}, #2 (#3)}
\def\phd#1#2#3{{ Physica D} {\bf #1}, #2 (#3)}

\def\prl#1#2#3{{ Phys. Rev. Lett.} {\bf #1}, #2 (#3)}

\def\pla#1#2#3{{ Phys. Lett. A } {\bf #1}, #2 (#3)}

\def\jcp#1#2#3{{J. Chem. Phys.} {\bf#1}, #2 (#3)}

\def\phd#1#2#3{{ Physica } {D #1} (#2) #3}
\def\prl#1#2#3{ Phys. Rev. Lett. { #1} (#2) #3}
\def\pla#1#2#3{ Phys. Lett.  { A #1} (#2) #3}
\def\pre#1#2#3{ Phys. Rev. E { #1} (#2) #3}

\def\jcp#1#2#3{{J. Chem. Phys.} { #1} (#2) #3}

\def\w{\omega}

\def\ad{amplitude death}

\def\le{Lyapunov exponent}
\def\ep{\epsilon}
\def\ie{i.e.~}
\def\eg{e.g.~}
\def\etl{$et ~al.$}

\def\ep{\epsilon}

\def\bc{\begin{center}}
\def\ec{\end{center}}
\def\eqn{\end{equation}\noindent}
\def\eqnr{\end{eqnarray}\noindent}
\def\beqr{\begin{eqnarray}}

\def\ep{\epsilon}
\def\eqn{\end{equation}\noindent}
\def\eqnr{\end{eqnarray}\noindent}
\def\beqr{\begin{eqnarray}}
 \def\beq{\begin{equation}}

\begin{document}

\begin{frontmatter}

\title{Amplitude Death:\\ The emergence of stationarity in coupled nonlinear systems}
\author[label1]{Garima Saxena}
\ead{gsaxena1@physics.du.ac.in}
\address[label1]{Department of Physics and Astrophysics, University of Delhi, Delhi 110007, India}
\author[label1,label2]{Awadhesh Prasad$^*$\footnote{$^*$Corresponding Author: Fax +11-2766 7061;
 Phone +11-2766 7725/Extn. 1354.}}
\ead{awadhesh@physics.du.ac.in}
\address[label2]{MPI-PKS,  N\"othnitzer Strasse 38, 01187 Dresden, Germany}
\author[label3,label4]{Ram Ramaswamy}
\ead{rr@uohyd.ernet.in}
\address[label3]{School of Physical Sciences, Jawaharlal Nehru University, New Delhi 110067, India} 
\address[label4]{University of Hyderabad,  Hyderabad 500046, India}

\begin{abstract}
When nonlinear dynamical systems are coupled, depending on the intrinsic dynamics and the manner in which the coupling is organized, a host of novel phenomena can arise.  In this context, an important  emergent phenomenon is the complete suppression of oscillations, formally 
termed {\it amplitude death} (AD). Oscillations of the entire system cease  as  a consequence of the  interaction, leading to stationary behavior. The fixed points that the coupling stabilizes can be  the otherwise unstable fixed points of  the uncoupled system or can correspond to 
novel stationary points. Such behaviour is of relevance in areas ranging from laser physics
to the dynamics of biological systems.    
In this review we discuss the characteristics of the different  coupling strategies and 
scenarios that lead to AD in a variety of different situations, and draw attention to 
several open issues and challenging problems for further study.

\end{abstract}


\end{frontmatter}

\tableofcontents

\newpage

\section{Introduction}
\label{intro}
The nature of the dynamics in coupled nonlinear systems is a central issue of interest in 
a number of areas of scientific enquiry.   A variety of natural phenomena--ranging from 
the motion of pendulums or springs, to lasers, fluid flow, ecological systems, populations,
and so on--are modeled as oscillators. These may be linear or nonlinear, depending on
the context, and as is well known, the motion can be periodic or chaotic. When 
systems that are in isolation capable of complex behaviour are coupled, 
a host of novel phenomena  can be seen. These  depend both upon the properties of the 
isolated systems---the nature of the nonlinearity, whether the motion is chaotic or regular,
what are the equilibria, and so on---as well as  
the manner in which their coupling is organized.  

An important {\it emergent} phenomenon  in this context is the suppression of oscillations,  
formally termed {\it amplitude death} (AD):  as  a consequence of the  interaction
oscillations of the entire system cease, leading to stationarity  \cite{eli1,turn,prig}. 
Such behaviour is emergent in the sense that the isolated or uncoupled systems do not exhibit 
stationary dynamics. In a formal sense,  amplitude death is an instance of a more general phenomenon that includes both the cessation of oscillations as a consequence of coupling,
as well the suppression of amplitude {\it variations}, namely the conversion of chaotic oscillations
to periodic or quasiperiodic dynamics. In a sense, this is the stabilization or creation
of ``simpler'' attractors through the coupling. 

When oscillations stop, there are two possibilities for AD.  If the coupled systems have 
exactly one equilibrium, then the occurrence of \ad~implies that this equilibrium becomes 
asymptotically stable. However, the coupled systems can have more than one 
stationary state and then one can have the stabilization of new fixed points, namely
those that are not stable (or may not even exist)  in the uncoupled system. 

A general setting in which we discuss AD in  this review is a configuration of
$N$ coupled nonlinear oscillators that is specified by the equation  
\begin{eqnarray}
\dot{\mathbf{X}}_i=\mathbf{F}_i(\mathbf{X}_i)+\epsilon \mathbf{G}_i(\mathbf{X})~~i=1,\ldots, N.
\label{eq:model}
\end{eqnarray} 
\noindent
where
\begin{eqnarray}
\mathbf{G}_i(\mathbf{X})=\frac{1}{K_i} \sum_{j = 1}^{N} A_{ij} \mathbf{H}(\mathbf{X}_i, \mathbf{X}_j,\tau),~~i=1,\ldots, N.
\label{eq:model1}
\end{eqnarray}
Here $\mathbf{X}_i$ is  the $m_i$-dimensional vector of dynamical variables for the $i$th oscillator,
and  the $\mathbf{F}_i$'s specify their evolution equations. The oscillators may be identical or 
distinct, and are coupled to other oscillators, as specified via the function $\mathbf{G}_i$. 
 $K_i$ is  a normalization factor, $\epsilon$ is the coupling strength, and 
the connection topology is specified by the connectivity matrix, whose  elements $A_{ij}$ are 1 or 0, depending on whether oscillators $i$ and $j$ are coupled with each other or not.  
The actual coupling is specified by the term $\mathbf{H}(\mathbf{X}_i, \mathbf{X}_j,\tau)$ which is a 
function  of $\mathbf{X}_i(t)$ and $\mathbf{X}_j(t-\tau)$. The time--delay $\tau$ can be 
fixed, or can have a more complex dependence on time.  Most of the major results discussed in this 
review are illustrated with two coupled oscillators, $N$ = 2.

\begin{figure}
\scalebox{0.4}{\includegraphics{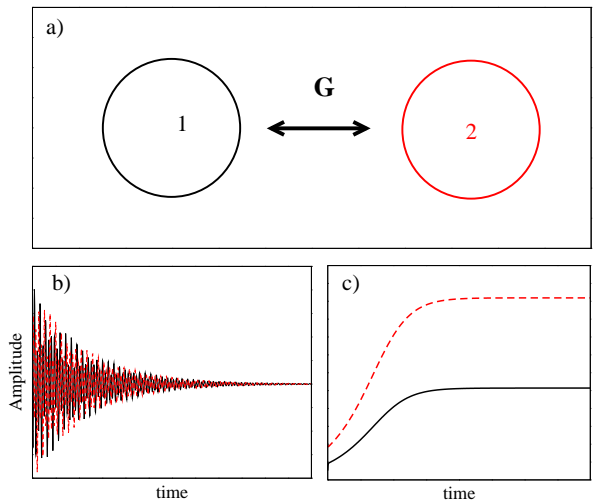}}
\caption{(Color online) Schematic  of the dynamics for a typical case of two oscillators with a given 
coupling function $\mathbf{G}$, showing (a) the phase space 
dynamics of the uncoupled oscillators, namely when $\epsilon$ = 0, and the transient
motions of the two subsystems (solid and dashed lines for respective oscillators) as a function of time,
 leading to either (b) the same (homogeneous) or (c) the distinct (heterogenous)
steady states respectively for appropriate coupling strength, $\epsilon \ne$0. }
\label{fig:model}
\vskip.1cm
\end{figure}

Shown in Fig. \ref{fig:model} is a schematic of the basic phenomenology. Consider two oscillators coupled by some 
function $\mathbf{G}$.  When the coupling strength $\epsilon= 0$ the motion of individual subsystems are 
oscillations (whether chaotic (C), periodic (P) or quasiperiodic (QP)); a simple periodic case is 
illustrated in Fig.~\ref{fig:model}(a). When the coupling is switched on, above a specific strength  when \ad~occurs,
 the transient dynamics is essentially that of damped oscillations, as shown in Figs. \ref{fig:model} (b) and (c).   
As can be seen, although this is an amplitude 
phenomenon, there could be additional effects that involve the phases of the oscillators,
and indeed, any analytic understanding of this phenomenon must include a discussion of
both the amplitudes and the phases of the interacting systems. 

There are some terminological issues that need to be clarified. In this article we will term the cessation of oscillations,
 regardless of what fixed points are stabilized asymptotically, as amplitude death, even though the asymptotic 
steady states in the different subsystems may not be the same  (as for example in Fig. \ref{fig:model}(c)) and
 can be different from the null state,  ${\bf X}^*_i \equiv$ 0. This heterogeneous case has been 
termed  ``oscillation death" \cite{eli1,kurth1,kurth2} or the Bar--Eli Effect \cite{aronson}), keeping the term 
amplitude death for the homogeneous case, when a simple transformation can shift the fixed point to the 
origin  \cite{aronson,reddy1} (as for example in Fig. \ref{fig:model}(b)). 
What these fixed points are will naturally depend on the internal structure of the individual oscillators, 
whether they are identical or mismatched, whether the coupling is symmetric or asymmetric, and indeed what the 
nature of the coupling is, namely whether the interactions  are diffusive or  nonlinear.

Following the initial work by Bar-Eli \cite{eli1}  AD has been studied quite extensively
 \cite{aronson,strogatz1,strogatz2,strogatz3}.  Different situations where \ad~occurs have been
described, and it is now known that the phenomenon can be seen when the interacting
systems  are identical \cite{reddy1,rajat1,rajatcc,atay1}, mismatched \cite{eli1,aronson}, when the coupling is dynamical  \cite{konishidc1}, or nonlinear (nondiffusive)  \cite{prasadnlc}. We review these various situations  and also discuss the general mechanism for AD in Sec. \ref{scn}. 

In many practical situations AD, is desirable, as for example in laser applications \cite{kim1,kim2,pramod1,pramod2,vishwa,prasad2} where a constant output is needed and fluctuations
should be suppressed. There are also other situations   where oscillations need to be 
maintained, as for instance in brain functioning \cite{leon1,leon2}. These different requirements suggest that control strategies to either to achieve or to avoid AD in coupled systems
are necessary, and in  Sec. \ref{target} we review some control methods that  have been
employed to either target or avoid steady states in coupled systems.

As indicated above, understanding the transient dynamics is important in describing AD since this
state is asymptotically featureless.  Transient dynamics can be significant in many applications that 
are restricted to finite times, as for example in ecology \cite{hastings}. Furthermore, since the different
subsystems are coupled, a study of  the synchronization properties \cite{piko} is also of interest. In a number
of examples it has been noted that the individual systems, while being synchronized, can 
nevertheless be either in--phase or out of phase, with a transition between these states at a specific 
value of the coupling \cite{prasad3,prasad4,appfb2}. We review this {\it phase--flip} transition and its associated behaviour in  Sec. \ref{character}. 

 The occurrence of  AD in networks of coupled oscillators  \cite{prasadnlc,rub,atay4,yang,erm1,wangn,chenn,wang}
is another topic of considerable current relevance. Studies have been carried out
for a variety of network topologies, ranging from systems coupled on a ring  \cite{sen,konishi10}, as well as in the so--called small world \cite{hou} and scale--free  networks \cite{liu}.  These are 
discussed in Sec. \ref{network}. 

Our principal aim in this review  on \ad~ is to stress the importance of this phenomenon and 
its applications in real situations.
The major  experimental results and possible applications 
are discussed in Sec. \ref{experiment}. Finally, the review concludes in  Sec. \ref{summary}  with 
a summary.

\section{Scenarios for Amplitude Death}
\label{scn}

To begin with, we consider the several known scenarios within which AD is known to occur, as 
well as constraints and necessary  conditions. One aspect of interest is in identifying the different types of interactions that facilitate AD. This is of particular relevance in natural systems where the parameter of individual systems may not be  accessible, and  \ad~ can be achieved mainly by  selecting appropriate forms of the interaction. It is also possible that for a given form of the interaction, AD occurs when internal parameters are varied. As a parallel, consider a related 
phenomenon, the synchronization of two systems. There are specific forms of interaction that cause
systems to synchronize, and for a specific form of coupling---diffusive, say---synchrony is possible only when the coupling strength is sufficiently strong. 

The Landau--Stuart limit cycle oscillator system \cite{aronson,landau,stuart} has been an
ideal model for the exploration of several of the scenarios we discuss here. This is a 
two--dimensional oscillator given by the evolution equation 
\begin{eqnarray}
\dot{Z}& =& [1+i\w-|Z|^2]Z,
\label{eq:ls}
\end{eqnarray}
where the complex variable $Z= x + i y$. As can be seen, the origin is an unstable fixed point, and the 
unit circle is an attractor for positive frequency, $\w$. The relative simplicity of the evolution equations
makes the analysis tractable, including the case of time--delayed interactions.  

In this section the case of two oscillators is considered, and  some numerical results
are presented first.  The stability of the fixed points can be estimated by computing the 
eigenvalues of the Jacobian matrix. When all eigenvalues have negative real parts,
the entire spectrum of Lyapunov exponents \cite{tabor} is negative: this is the state
of AD.  In the numerical results presented here the standard Runge--Kutta scheme 
\cite{press}  was used. 

\subsection{Mismatched oscillators}
\label{mismatch}

\begin{figure}
\scalebox{.5}{\includegraphics{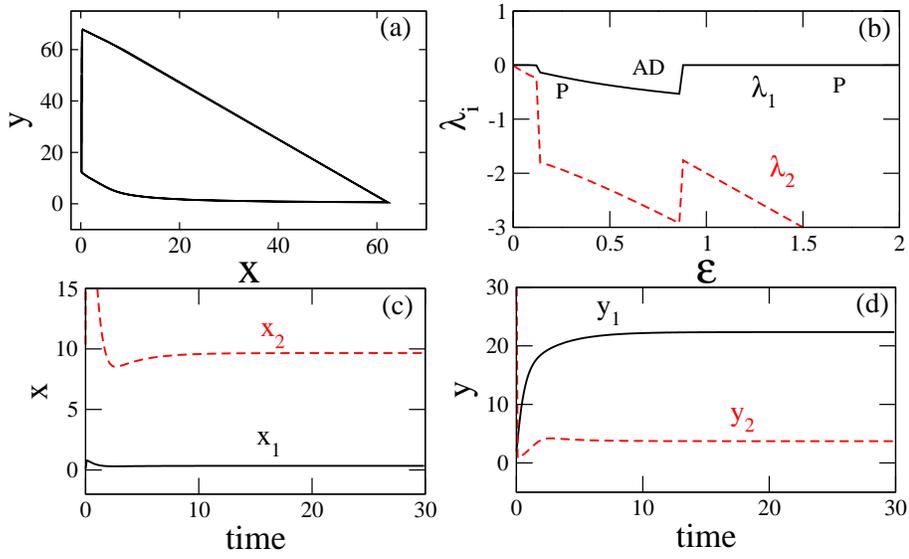} }
\caption{ (Color online) (a) The trajectory of a single Brusselator, Eq. (\ref{eq:bruss}), at 
parameters $A=5$ and $B=35$ in phase space.
(b) The largest two Lyapunov exponents  as a function of coupling strength $\ep$  for
two coupled identical Brusselators, Eq. (\ref{eq:eli}) at parameters $A_1=A_2=5$ and $B_1=B_2=35$.
(c) And (d) show the transient trajectories of both oscillators as function of  time at coupling strength $\ep=0.5$.}
\label{fig:bru}
\end{figure}

An early observation  of amplitude quenching was reported by Crowley and Field \cite{crowley1} who experimentally coupled two chemical oscillators, the Belousov--Zhabotinski system electrically. 
The coupling essentially involved the mass transfer of a single species, the Ce$^{3+}$ ions, and caused the composite system to reach a state wherein one of the systems was quenched to a steady state,  while the other maintained oscillations.  This experiment served as the starting point for the study of this behaviour by  Bar-Eli \cite{eli1}  who showed  that two interacting model continuous stirred tank reactors can stop oscillations and arrive at steady states if coupled diffusively. 
Consider the cell Brusselator model \cite{eli1},  defined as 
\begin{eqnarray}
\nonumber 
\dot{x} &=&-(B+1)x+x^{2}y+A \\
\dot{y}&=&Bx-x^{2}y 
\label{eq:bruss}
\end{eqnarray} 
where  $x$ and $y$ are the dimensionless concentrations of two chemical species \cite{prig,eli1} while the (positive) parameters $A$ and $B$ are the rate constants for the production of $x$ and $y$. 
This system has  one steady state, $x^{*}=A, y^{*}={B}/{A}$ which is unstable if $A^2<(B-1)$.
The oscillating behavior of this system is shown in Fig. \ref{fig:bru}(a) for parameters $A=5$ and $B=35$. When two such cells are coupled diffusively \cite{eli1} then the dynamical equations
take the form (the variables of the two systems are distinguished by subscripts)
\begin{eqnarray}
\dot{x_1}&=&-(B_1+1)x_1+x_{1}^{2}y_1+A_1 +\ep (x_2-x_1)\nonumber \\
\dot{y_1} &=&B_1 x_1-x_1^{2}y_1 +\ep (y_2-y_1) 
\label{eq:eli}
\end{eqnarray} 
\noindent
with corresponding equations for subsystem 2. The coupling terms $\ep (x_2-x_1)$ and $\ep (y_2-y_1)$ arise from material transfer between the cells through diffusion, and the coupling strength
 $\ep$ corresponds to the rate of this transfer.  The largest two Lyapunov exponents at $A_1=A_2=5, B_1=B_2=35$ as a function of the coupling strength  in Fig.~\ref{fig:bru}(b). 
The transition to AD can be identified as the point when all  \le s become nonpositive (and which then correspond 
to the real parts of the eigenvalues of the fixed point--see Sec. \ref{character}). 
Both the oscillators then settle to different  steady states.
 
Transient trajectories for both oscillators are  shown in Figs. \ref{fig:bru}(c) and (d), demonstrating that this is a case of oscillation death: the two fixed points ($x^*_{1}=0.34, y^*_{1}=22.36)$ and ($x^*_{2}=9.66, y^*_{2}=3.72)$ are distinct, and different from the fixed point of the uncoupled systems, namely $x^*= 5, y^* = 7$.  

Similar behavior was found to exist in other well--known chemical oscillators \cite{eli1,eli4},
including the Noyes--Field--Thompson model, the Oregonator,  and the Field--k\"{o}r\"{o}s--Noyes models for the  Belousov--Zhabotinski reaction, the model of first-order decomposition autocatalysis \cite{kumar}  and  the Lotka--Volterra model for prey--predator interactions. These systems differ significantly showing that the occurrence of \ad~ is fairly general. This extensive  numerical analysis \cite{eli1}  triggered a serious exploration of the phenomenon of \ad~from both a theoretical and experimental point of view. 
 
\begin{figure}
\scalebox{1.}{\includegraphics{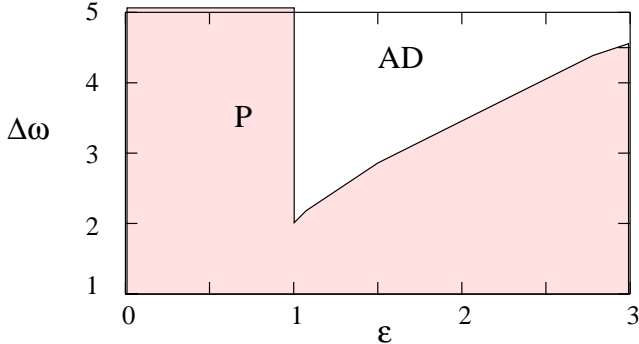} }
\caption{ The phase diagram in the parameters: frequency mismatch $(\Delta \omega)$ and coupling strength $(\ep)$ of
 Eq. (\ref{eq:ls-ins}).
The shaded  and unshaded regions correspond to the periodic (P) and \ad~(AD) motions respectively.
For detail see Ref. \cite{aronson}.}
\label{fig:aronson}
\end{figure}

\begin{figure}
\scalebox{.5}{\includegraphics{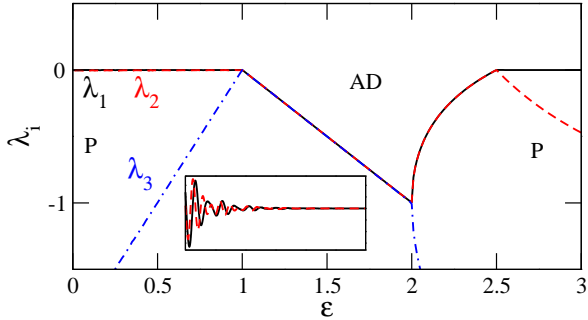} }
\caption{(Color online) The largest two Lyapunov exponents as a function of coupling strength,
 at $\Delta\omega=4$ ($\omega_1=4$ and $\omega_2=8)$, 
Eq. (\ref{eq:ls-ins}). Inset figure shows in--phase synchronous behaviour of transient trajectories of both the 
oscillators as a function of time at coupling strength $\ep=1.5$ , $x_1$ (solid-black line)
 and $x_2$ (dashed-red line). }
\label{fig:aronson1}
\end{figure}

Analysis by Aronson, Ermentrout, and Kopell \cite{aronson} provided a deeper mathematical understanding of this phenomenon. They considered the following system  of two Landau--Stuart oscillators, 
\begin{eqnarray}
\dot{Z}_{1}&=&[1+i\w_{1}-|Z_{1}|^2]Z_{1}+\epsilon[ Z_{2} -Z_{1}],\nonumber\\
\dot{Z}_{2}&=&[1+i\w_{2}-|Z_{2}|^2]Z_{2}+\epsilon[ Z_{1} -Z_{2}]
\label{eq:ls-ins}
\end{eqnarray}
\noindent 
where the notation is obvious.  When uncoupled, both oscillators have an
unstable fixed point at $Z^*_{1,2}=0$, and the unit circles $|Z_{1,2}|=1$ are
limit cycles.
Depending on the mismatch $\Delta \omega =|\omega_1-\omega_2|$, this system shows different
types of dynamics  \cite{aronson}. Shown in Fig. \ref{fig:aronson} is a schematic phase--diagram, indicating the different dynamical states:  the shaded and blank regions correspond to 
periodic motion and AD  respectively.  For a specific value of the 
mismatch, $\Delta \omega = 4$, the largest two  \le s are shown in Fig. \ref{fig:aronson1}, as a function
of the coupling. The transient dynamics in the AD region is shown in the inset; note that this is a case of homogenous steady states and both oscillators settle 
onto the (now stabilized) fixed point, the origin $Z^*_{i}=0$.  Linear stability analysis in the neighborhood of the origin shows that  \ad~ occurs for  $\ep >$ 1 and $\Delta \omega>2\sqrt{2\epsilon-1}$.

The sign of $\w$ determines the sense of rotation in phase.  In Eq. (\ref{eq:ls-ins}) the 
frequencies are both positive so that in the phase plane the direction of motion for both oscillators
is the same. In this case, the systems are termed co--rotating, while when the frequencies differ in sign, they are counter--rotating \cite{prasad1}. This latter situation is one of parameter mismatch, and 
in recent works \cite{prasad1,dana,manish2}, the characteristics of AD in such cases have been
discussed; see also Sec. \ref{character}.

Parameter mismatch is thus one of the major causes of \ad, and has been extensively studied both analytically and experimentally \cite{kurth1,kurth2,aronson,strogatz2,erm1,prasad1}.

\subsection{Delay interaction}
\label{delay}
In the above systems, the interaction is considered to be instantaneous, namely the coupling terms involve the variables of the two subsystems at the same time, $t$. In many physical systems, the coupling involves time--delay, and this was considered by Reddy, Sen and Johnston \cite{reddy1}
who investigated the collective dynamical behavior of limit--cycle oscillators 
interacting diffusively via time--delayed coupling.  Here AD was achieved even in 
the absence of mismatch, and was demonstrated in an experiment involving electronic
circuits.  As was shown through both analysis and numerics \cite{reddy1,reddy2,reddy3,reddy4,senthil},
there is a significant region in the parameter space of coupling strength and time delay
where AD occurs. 

The delay--coupled Landau--Stuart system has the evolution equations
\begin{eqnarray}
\dot{Z}_{1}&=&[1+i\w_{1}-|Z_{1}|^2]Z_{1}+\epsilon[ Z_{2}(t-\tau) -Z_{1}],\nonumber\\
\dot{Z}_{2}&=&[1+i\w_{2}-|Z_{2}|^2]Z_{2}+\epsilon[ Z_{1}(t-\tau) -Z_{2}],
\label{eq:reddy}
\end{eqnarray} \noindent
where the output of one system takes a time $\tau$ to influence the other, as might occur
in spatially separated systems when there is a finite speed of signal transmission. 

\begin{figure}
\scalebox{.5}{\includegraphics{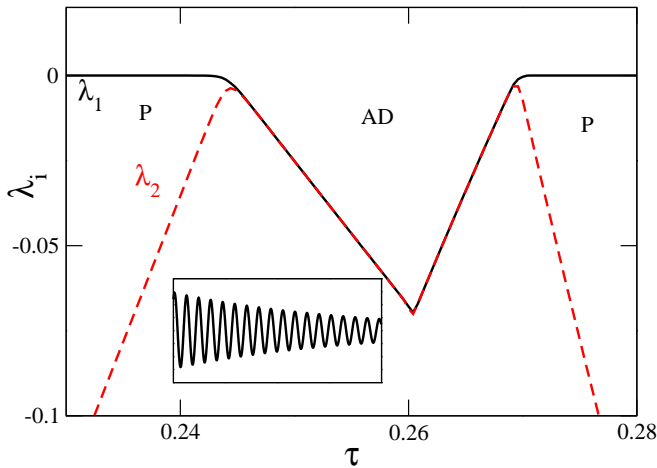} }
\caption{ (Color online) The largest two Lyapunov exponents as a function of coupling strength for coupled
identical oscillators for $\Delta\omega=0$ ($\omega_1=6$ and $\omega_2=6)$,
 (Eq. (\ref{eq:reddy})). Inset figure: completely synchronized transient trajectories, $x_1$ (solid-black line)
 and $x_2$ (dashed-red line), as a function of time at time-delay $\tau=0.255$ \cite{reddy1,prasad3}.}
\label{fig:reddy1}
\end{figure}

Consider identical systems,  $\omega_1=\omega_2$ = 6. Shown in Fig. \ref{fig:reddy1} 
is the largest two \le s as a function now of the time-delay $\tau$ for fixed coupling strength $\ep= 2$. This clearly shows the regime of AD when all \le s are negative (see details in Sec. \ref{route}). Transient trajectories (shown in the inset) for both oscillators indicates the cessation of oscillations
and the stabilization of the homogeneous steady state, $Z^*_{1,2}=0$. The AD in such delayed coupled systems has a resonance--like character, occurring in repeated ``islands" in the 
parameter space of $\ep$ and $\tau$ \cite{reddy1,reddy2,reddy3,reddy4,senthil}. 

This scenario for the occurrence of AD in delay--coupled systems is quite general and does not
depend on the nature of the dynamics in the uncoupled systems. As was shown by Prasad \cite{prasad3}, even chaotic oscillators can be stabilized in this manner. The importance of time delay  interactions has been highlighted in \cite{strogatz5}. 

Given a finite transmission speed for signals from one subsystem to reach the other, a natural extension is to consider the situation when the time--delay itself varies. This is particularly important
when there are stochastic effects that are being modeled by the delay itself. Atay \cite{atay1} considered the effect of a distribution of time--delays and showed that even a small spread in the delay distribution can greatly enlarge the set of parameters for which  amplitude death occurs.  

The expansion of the AD region with a distribution of delays is a important effect with a wide range of applications \cite{atay1}, particularly since if the variance of the distribution is above a threshold, this AD region can become unbounded.  Again, the effect is quite general, occurring for arbitrary spread in the delay, for different forms of the distribution,  and for arbitrary numbers of oscillators \cite{atay1,atay2}. 
Closely related to the case of distributed delays is the situation where systems respond to 
cumulative signals, namely by integrating information received over an interval in time. This occurs
when systems  have a finite intrinsic response time and also causes the region of AD to extend
indefinitely  \cite{garima1,garima2}. 

Since time--delay interactions are common in experiments, various aspects
of both fixed and distributed delays, including a deterministic time-varying interaction \cite{michiels,gjurchinovski} have been studied. One application is to 
the  stabilization of fixed points \cite{konishi3,konishi4,konishi5}. Velocity delayed coupling has been also explored in the context of  \ad~\cite{song,lia} (see Sec. \ref{sec:velocity}). Mixed interaction schema, with instantaneous 
interactions in one system and time--delay in the other \cite{zou} or unequal time--delays in different directions (to model spatial heterogeneity  \cite{sun,rajat3} and one-way ring time-delay \cite{konishi10} have been studied in recent works.  


\subsection{Conjugate coupling}
\label{conjugate}

A novel context for \ad~ is when systems are coupled via so--called conjugate or dissimilar 
variables. With such coupling, the necessity for systems being either mismatched or having
time--delayed interactions can be dropped, and AD occurs in identical systems with 
instantaneous coupling. This was first studied in a model system by Karnatak \etl~\cite{rajat1} 
who  considered two Landau--Stuart oscillators with the evolution equations 
\begin{eqnarray}
\nonumber
\dot{x}_{1}&=&P_1x_1-\omega_1y_1+\ep(y_2-x_1)\\
\dot{y}_{1}&=&P_1y_1+\omega_1x_1+\ep(x_2-y_1)
\label{eq:rajat}
\end{eqnarray} \noindent
where $P_i=1-|Z_i|^2$, and similar equations can be written for subsystem 2. Note that here the $y$ variable of the second oscillator is diffusively coupled to the $x$ variable of the first oscillator 
and vice versa. 

\begin{figure}
\scalebox{.5}{\includegraphics{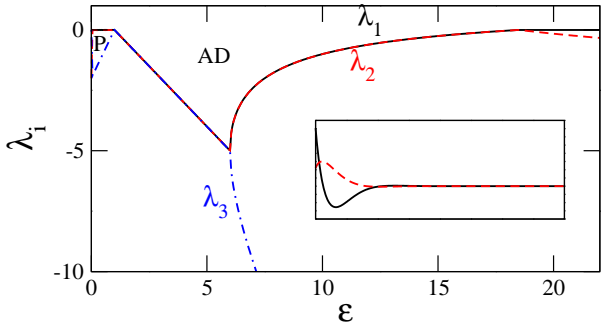} }
\caption{ (Color online)  The largest three  Lyapunov exponents as a function of coupling strength at $\Delta\omega=0$ ($\omega_1=6$ and $\omega_2=6)$, 
Eq. (\ref{eq:rajat}).
 Inset figure: the transient trajectories, $x_1$ (solid-black line) 
 and $x_2$ (dashed-red line), as a function of time at coupling strength $\ep=5$ \cite{rajat1}.}
\label{fig:rajat1}
\end{figure}

Coupling via conjugate variables is natural in a variety of experimental situations, and indeed the
above system was inspired by  the experiments of Kim and Roy \cite{kim1,kim2} on coupled semiconductor laser systems  where the photon intensity fluctuations from one laser were used to 
modulate the injection current of a second identical laser and vice versa. Indeed it would not
be possible to couple two lasers through their intensity fluctuations directly. As it happens, in a
model of the semiconductor laser, the current and the fluctuations are, in a sense, conjugate
variables that have constitute the chaotic oscillator system \cite{kim1,kim2}.
The largest three Lyapunov exponents for the above system is shown in Fig. \ref{fig:rajat1}, and as can be seen, there is a wide range of coupling strength where all 
 \le s are negative, indicating stabilization of the fixed point. A transient trajectories are
shown in the inset.

A detailed analytical condition for the occurrence of \ad~ can be found in \cite{zhang}. With 
conjugate coupling, AD can be anticipated by recognizing that there is a strong analogy
between time--delayed and conjugate variables: as a matter of fact, the Takens'  
embedding theorem \cite{taken} asserts that the topological properties of a dynamical system that is reconstructed through delay variables match those of the true system for appropriate choices of embedding dimension and time delay. Thus using conjugate variables is akin to using time--delay
coupling, and this gives rise to regimes of AD. 
 
In addition, there are other new features such as the suppression of chaos, and riddling
 \cite{rajat1,rajatcc,rajat2} due to conjugate coupling.  In addition, this is particularly 
 suited to experiments and both in electrochemical systems \cite{dasgupta} as well as
 in  electronic circuits \cite{singla}, this coupling can be realized.

\subsection{Dynamic coupling}
\label{dynamical}

\begin{figure}
\scalebox{.5}{\includegraphics{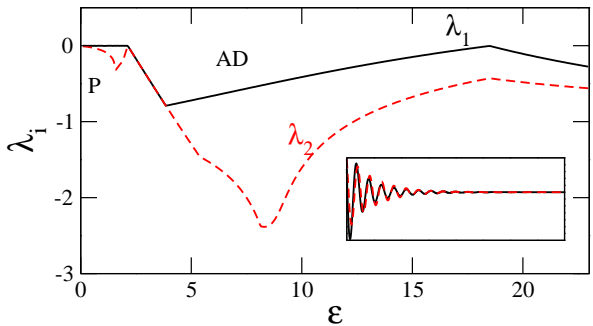} }
\caption{ (Color online) The largest two Lyapunov exponents  as a function of coupling strength, at
 $\Delta\omega=0$ ($\omega_1=6$ and $\omega_2=6)$, Eq. (\ref{eq:konishi}). Inset figure: transient trajectories,
$x_1$ (solid-black line) and $x_2$ (dashed-red line), as a function of time  at coupling strength $\ep=3$ \cite{konishidc1}.}
\label{fig:konishi1}
\end{figure}

Konishi \cite{konishidc1} has proposed another type of interaction which,  
in contrast to the cases discussed above
has an evolving or ``dynamic'' coupling.  Consider the Landau--Stuart system 
\begin{eqnarray}
\dot{x}_{1}&=&P_1x_1-\omega_1y_1+\ep(u_1-x_1)\nonumber\\
\dot{y}_{1}&=&P_1y_1+\omega_1x_1\nonumber\\
\dot{u}_{1}&=&-u_1+x_2
\label{eq:konishi}
\end{eqnarray}
with similar equations for subsystem 2, and where $P_i=1-|Z_i|^2$ as usual. The variables
$u_i$ follow linear dynamics in absence of the $x_j$'s  so that the coupling has its own
nontrivial dynamics.  Shown in Fig. \ref{fig:konishi1} is the largest two Lyapunov exponents;
the occurrence of \ad~ is evident, and the transient dynamics  are shown in the inset. 
Note that this instance of \ad~ is in identical systems.  
In addition to estimating necessary and sufficient conditions
for amplitude death to occur, this form of the dynamics has 
lent itself to extensive analytic study \cite{konishidc2,konishidc3,konishidc4}.


\subsection{Nonlinear Coupling}
\label{nonlinear}

In the preceding examples, the form of the coupling has been taken to be linear or diffusive. An interesting extension that uses nonlinear coupling makes it possible to not only achieve 
AD, but also to specify the steady state of the coupled system  \cite{prasadnlc,prasad2}. We
will illustrate the targeting of specific fixed points  in Sec. \ref{nonlinear-target}.
Consider again the Landau--Stuart oscillators
\begin{eqnarray}
\dot{x}_{1}&=&P_1x_1-\omega_1y_1+\ep(x_1-\alpha)\exp(x_2-\beta) \nonumber\\
\dot{y}_{1}&=&P_1y_1+\omega_1x_1.
\label{eq:nlc}
\end{eqnarray}\noindent
where the coupling is nonlinear (and similar equations hold for subsystem 2). The motivation for this choice of function is discussed in Sec. \ref{nonlinear-target}, and is common in neuronal systems \cite{prasadnlc,prasad2}.

\begin{figure}
\scalebox{.5}{\includegraphics{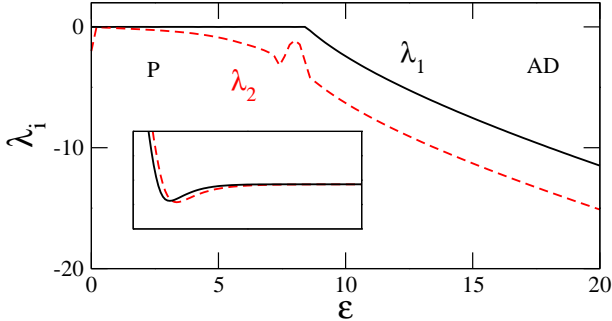} }
\caption{ (Color online) The largest two Lyapunov exponents as a function of coupling strength, at 
$\Delta\omega=0$ ($\omega_1=6$ and $\omega_2=6$, $\alpha=1$) and $\beta=0.1$, Eq. (\ref{eq:nlc}).
Inset figure: the transient trajectories, $x_1$ (solid-black line) 
 and $x_2$ (dashed-red line), as a function of time at coupling strength   $\ep=15$ \cite{prasadnlc,prasad2}.}
\label{fig:nonlinear1}
\end{figure}

Shown in  Fig. \ref{fig:nonlinear1} is the largest two Lyapunov exponents for identical oscillators as a function of $\ep$, the region where all \le s are negative corresponds to \ad~ 
(the transients are shown in the inset). We have found that the 
phenomenon of \ad~ when systems are coupled through nonlinear interactions is very general
and occurs in the absence of parameter mismatch, in the absence of time--delay  (although time delay can enhance the effect)  \cite{prasadnlc,prasad7}.

\subsection{Linear augmentation, and other strategies}
\label{manish}
\begin{figure}
\scalebox{.5}{\includegraphics{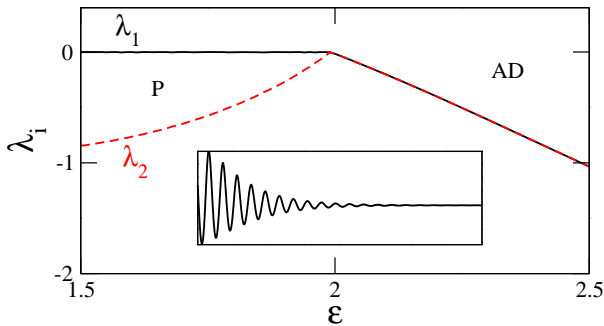} }
\caption{ (Color online) The largest  two Lyapunov exponents as a function of coupling
 strength at $\Delta\omega=0$ ($\omega_1=6$ and $\omega_2=6$, $K=1$) and $\beta=1$, Eq. (\ref{eq:linear}).
 Inset figure: transient trajectory, $x$, as a function of time at coupling strength $\ep=2$ \cite{manishla}.}
\label{fig:manish1}
\end{figure}

A final scenario we discuss dispenses with the need that the interacting oscillators be similar. 
Recently Sharma \etl~ \cite{manishla} have proposed a new strategy through which a nonlinear oscillator, when coupled to a linear system,  experiences \ad. Consider such a 
Landau--Stuart oscillator
\begin{eqnarray}
\dot{x}&=&Px-\omega y+\ep(u-x)\nonumber\\
\dot{y}&=&Py+\omega x\nonumber\\
\dot{u} &=&-k u-\ep(x-\beta).
\label{eq:linear}
\end{eqnarray}
\noindent
where $\epsilon$ is the coupling strength.  The variable $u$ describes the dynamics of the linear system with decay parameter $k$ (details are given in Sec. \ref{la}), and the largest two Lyapunov
exponents in Fig. \ref{fig:manish1}, clearly shows the region of   \ad. 

Such interaction has been termed linear augmentation, in the sense 
that the dimension of the oscillator is  increased by one through the addition of the 
linear equation. This interaction is also dynamical  (cf. \ref{dynamical}), 
the main difference between the two strategies, Eqs. (\ref{eq:konishi}) and (\ref{eq:linear})
is that the interaction of each oscillator is via a  linear evolution equation in the former case,
while a single oscillator is used as self--feedback  in the latter case. 
Such forms of interaction, Eq. (\ref{eq:linear}), are useful in the context of  targeting unstable fixed points,  as shown in Sec. \ref{la}.

Apart from above discussed scenarios, other situations where  \ad~ occurs include the case of 
indirect coupling \cite{manish2,resmi1,manish3}: when two oscillators are coupled via a third, 
the presence of the intermediate system causes an effective  ``transmission'' delay, which then
effects \ad. It has been also suggested \cite{resmi1} that the AD is due to competition between  synchronization  and anti synchronization. Similarly, the conflict between attractive and 
repulsive (\ie negative and positive) diffusive coupling  also gives rise to AD \cite{chen}. 
AD has also been seen in parametrically modulated systems \cite{pisa}, phase repulsive communication \cite{ullner}, and by forcing \cite{lin} or gradient coupling \cite{zou}.

\subsection{Velocity Coupling}
\label{sec:velocity}
The discussion above has focused entirely on coupled dissipative systems. 
The Lyapunov spectrum of conservative Hamiltonian systems 
is constrained by symmetry considerations to have an equal number of 
positive and negative exponents and to sum to zero. Thus these 
cannot show AD. However, when two conservative systems are coupled 
by time--delayed velocity coupling these can show AD \cite{garima3}.

Consider, for instance two non-integrable H\'enon-Heiles systems
\cite{tabor} with time--delayed velocity coupling as
\begin{eqnarray}
\nonumber
\ddot{x}_{1}&=&-x_1-2x_1y_1+\epsilon (\dot{x}_2(t-\tau)-\dot{x}_1(t))\\
\nonumber
\ddot{y}_{1}&=&-y_1-x_1^2+y_1^2\\
\nonumber
\ddot{x}_{2}&=&-x_2-2x_2y_2+\epsilon (\dot{x}_1(t-\tau)-\dot{x}_2(t))\\
\ddot{y}_{2}&=&-y_2-x_2^2+y_2^2.
\label{eq:hh}
\end{eqnarray}
\noindent 

\begin{figure}
\scalebox{.5}{\includegraphics{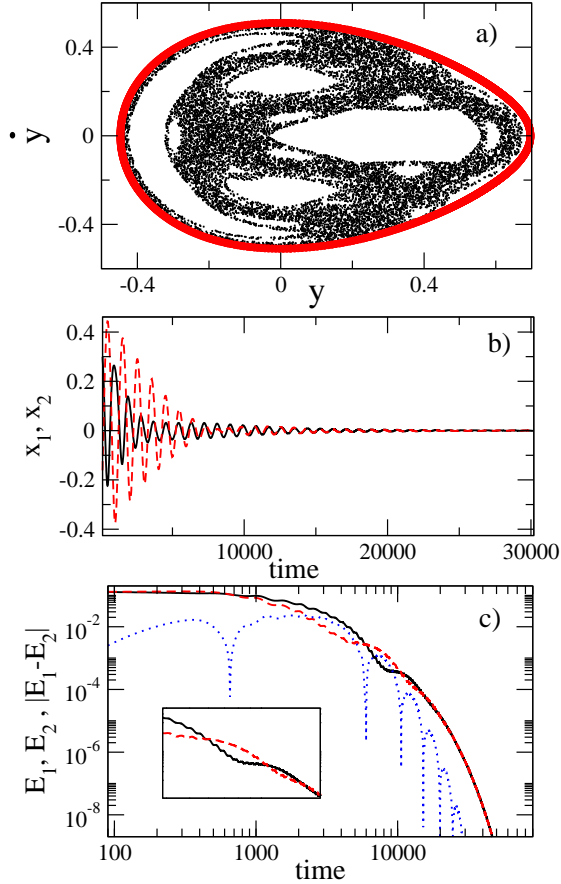} }
\caption{ (Color online) (a) Poincare map at $x=0$ of uncoupled conservative Henon-Heiles system, Eq. (\ref{eq:hh}) 
 at energy $E=0.13$. Outer red points and inner black points correspond to the regular and chaotic motions respectively.
 (b) The transient trajectories of $x_1$ (solid-black line)
and $x_2$ (dashed-red line) as a function of time at coupling strength $\epsilon=0.1$ and time-delay $\tau=2$.
 (c) The  dissipation of energy $E_1$ (solid-black line), $E_2$ (dashed-red line)
of individual oscillators and their energy difference $|E_1-E_2|$ (dotted-blue line) as a function of time.
 Inset figure show the repeated crossing of energies while decaying (Ref.  \cite{garima3}).}
\label{fig:hh}
\end{figure}

As is well--known, in the uncoupled case namely for $\epsilon=0$ the dynamics has both regular and irregular behavior largely depending on the total energy $E_i$ although there is also considerable initial condition dependence \cite{tabor}. The individual energies are 
\beq E_i= \frac{\dot{x_i}^2+\dot{y_i}^2}{2}+\frac{{x_i}^2+{y_i}^2}{2}+x_i^2y_i-\frac{y_i^3}{3}.
\eqn 
Shown in Fig. \ref{fig:hh}(a) are  the Poincar\'e maps for two different initial conditions, one leading to 
regular motion (outer points), while one leads to chaotic dynamics (inner points), at the same energy $E$ = 0.13, just below the dissociation limit, $E$ = 1/6.  

When velocity coupling is introduced say for  $\epsilon=0.1$ and $\tau=2$, the system effectively becomes dissipative and the dynamics is attracted to the origin. Transient trajectories are shown in Fig. \ref{fig:hh}(b). The loss of energy has been discussed in a related context by Wang et. al \cite{wang1} who showed that in the neighborhood of the fixed point,  \ad~ occurs when the 
averaged total power is negative definite. In the case of the coupled H\'enon--Heiles system 
the energy loss of the individual oscillators, $E_i$ and their difference $|E_1-E_2|$ are shown in
Fig. \ref{fig:hh}(c)  \cite{garima3}).  
The question of whether other fixed points with nonzero asymptotic energies can be targeted
so that the amplitude death reaches a nontrivial steady state remains open.

The above  examples demonstrate that with a proper choice for the interaction, a variety of systems
can show AD, thus making such steady states amenable to control. 

\subsection{Routes to AD}
\label{route}
In this section we consider the mechanisms for \ad, namely the ``routes'' followed as parameters in the system are varied. Since the transition to AD is from oscillatory motion  to a fixed point, an immediate question is whether there is a bifurcation to AD \cite{suarez}, in analogy with other standard routes 
that occur in nonlinear dynamical systems \cite{ott,strogatz-book,kuz,kaneko}.

\subsubsection{Hopf bifurcation}
\begin{figure}
\scalebox{.5}{\includegraphics{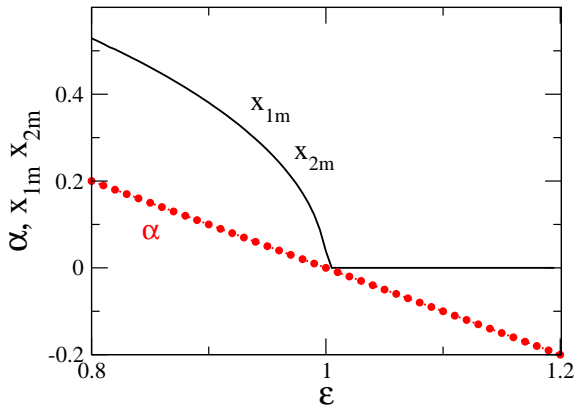} }
\caption{ (Color online) Maxima of $x_1$ and $x_2$ (overlapped solid lines) and real part of the 
largest eigenvalues (dashed line with circles) as a function of  coupling strength, 
$\ep$; see Eq. (\ref{eq:ls-ins}).}
\label{fig:hb}
\end{figure}

The most widely observed route to AD is through the Hopf bifurcation,  where the coupling induces stability of the fixed point of the uncoupled systems. In the Landau--Stuart oscillators, Eq. (\ref{eq:ls-ins}), the uncoupled system has two attractors, namely the stable limit cycle and an unstable fixed point.
At the transition that occurs at $\epsilon$ =1 (see Fig. \ref{fig:aronson1}) the 
  real part of  the largest eigenvalue at the origin, $Z^*_{i}$ = 0 becomes negative: this is the standard Hopf bifurcation since in fact a pair of eigenvalues cross the axis from right to left
in the complex plane \cite{strogatz-book}. The amplitudes also go to zero at the same point, namely $\ep$ = 1, as shown in Fig. \ref{fig:hb} for different frequencies, $\w_1=4$ and $\w_2=8$. 
  
\subsubsection{Saddle-node Bifurcation}
\label{multi}

Another route to AD when {\it new} fixed points are created is via a saddle--node bifurcation. 
These new fixed points annihilate the periodic orbit, thereby causing the oscillations to stop.

\begin{figure}
\noindent
\scalebox{.5}{\includegraphics{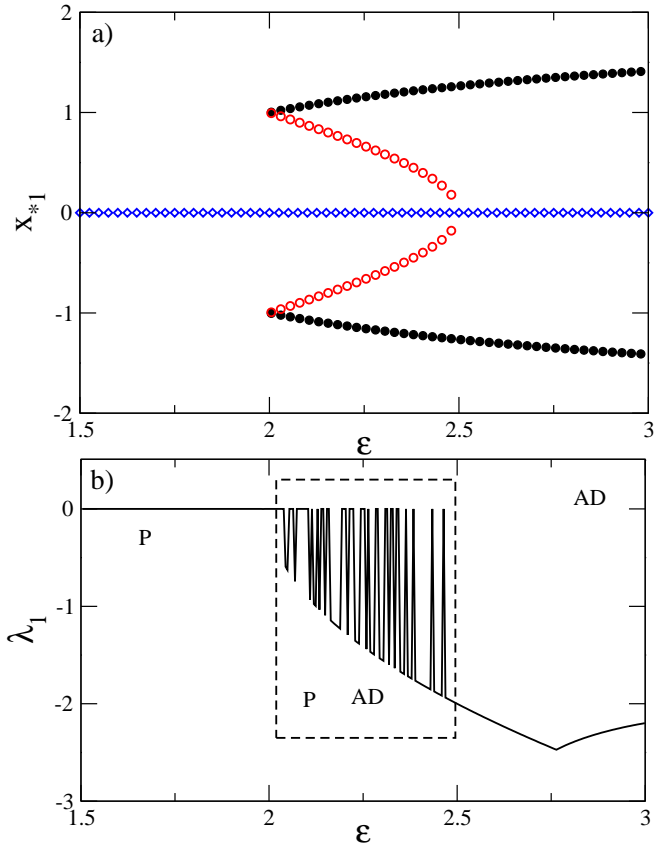} }
\caption{ (Color online) (a) The $x*_{1}$ component of   fixed points: trivial (blue open diamonds), nontrivial unstable (red open circles) and 
stable  (black filled circles) with coupling strength $\epsilon$ for conjugately one way coupled limit cycle 
oscillators, Eq. (\ref{eq:multi}), for $\omega=2$.
(b) The largest  Lyapunov exponent  as a function of coupling strength.
Marked box shows the multistable region  where periodic as well as AD co-exist \cite{rajat1,suarez}.}
\label{fig:multi}
\end{figure}

For illustration consider conjugate coupled Landau--Stuart oscillators (cf. Eq. (\ref{eq:rajat}))
\begin{eqnarray}
\nonumber
\dot{x}_{1}&=&P_1x_1-\omega_1y_1\\
\dot{y}_{1}&=&P_1y_1+\omega_1x_1+\ep x_2.
\label{eq:multi}
\end{eqnarray} \noindent
The coupling, which is nondiffusive results in the creation of new a fixed point \cite{rajat1},  
in addition to the origin (0,0,0,0), at ($x^*_{1},y^*_{1},x^*_{2},y^*_{2}$) where
\begin{eqnarray}
x^*_{1}&=&\pm \sqrt{(\omega/\epsilon (1\pm\sqrt{\epsilon\omega-\omega^{2}}} \nonumber \\
y^*_{1}&=&\pm \sqrt{1-(x^*_1)^2\pm\sqrt{\epsilon\omega-\omega^{2}}},
\label{eq:multi-x}
\end{eqnarray}
with  $x^*_{2}=-x^*_{1}$ and $y^*_{2}=-y^*_{1}$.
Note that this second fixed point depends on the coupling strength, and exits 
only when $\epsilon >\omega$ 
(see the filled and open circles in Fig. \ref{fig:multi}(a)). 
The trivial fixed point  (0,0,0,0)  (open diamond in Fig. \ref{fig:multi}(b)) which  exists for all values of coupling strengths is always unstable, but the stability of the nontrivial fixed point changes with 
coupling. The stable and unstable points collide at $\epsilon = \omega$, causing the periodic motion: 
thus \ad can be achieved with such a  saddle--node bifurcation \cite{suarez}. Note also that
the oscillators settle on different fixed points that depend upon initial conditions and thus this is 
an instance of  inhomogeneous amplitude death. This route to AD has been found 
in other systems as well \cite{suarez}. 

The largest Lyapunov exponent, shown in Fig. \ref{fig:multi}(b)  shows strong fluctuations in the region
prior to the AD transition, namely the marked box
$\epsilon\in [\omega, (1+\omega^2)/\omega]$ suggesting that  there is multistability with both periodic motion, ($\lambda_1$ = 0)  and AD ($\lambda_1<0$) coexisting. Details are given in \cite{rajat1}.
This indicates that apart from the nontrivial fixed point solutions there is also a periodic solution.
Multistability with a riddled basin has been seen in diffusively coupled R\"ossler chaotic  oscillators \cite{rajat1} as well as dissimilar oscillators (R\"ossler and Lorenz systems)  \cite{prasad3,chen}. In the riddled region, vanishingly small changes in initial conditions lead to different attractors, making
the system completely unpredictable \cite{ott,alex}. This show that near the onset of AD there is the possibility of complex  dynamics \cite{rajat1,prasad3,suarez,prasad6} as well.

\subsubsection{Direct transition}

\begin{figure}
\scalebox{.5}{\includegraphics{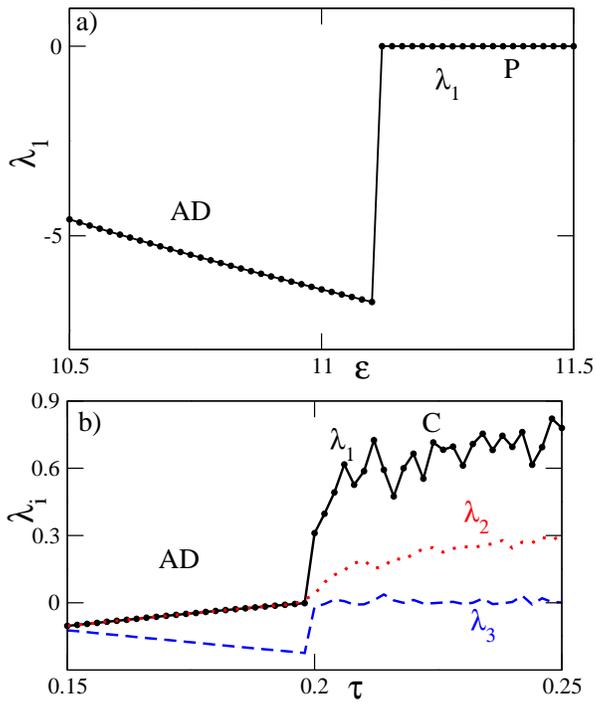} }
\caption{ (Color online) The largest Lyapunov exponent as a function of coupling strength for coupled limit cycle oscillators, Eq. (\ref{eq:ls-ins})
\cite{prasadnlc}, (b) The largest three  Lyapunov exponents as a function of time-delay for coupled chaotic oscillators Eq. (\ref{eq:lorenz}) at fixed coupling
 strength $\ep= 0.5$ \cite{prasad3}. }
\label{fig:direct}
\end{figure}

Apart from the Hopf and saddle--node bifurcations, a third type of transition to  AD
has been reported. Consider the system, Eq. (\ref{eq:nlc}) where the interaction 
between the subsystems is nonlinear.  Shown in Fig. \ref{fig:direct}(a) is
the largest \le~ at higher value of coupling (extension of Fig. \ref{fig:nonlinear1}):
there is a jump from periodic motion ($\lambda_1$ = 0) to AD ($\lambda_1 < $0). 
This abrupt change in the dynamics could be a bifurcation but has not been analyzed 
completely. Similar transitions have also been observed in coupled chaotic
oscillators as for instance two Lorenz oscillators with time-delay 
coupling  \cite{prasad3}
\begin{eqnarray}
\nonumber
 d x_1(t)\over dt &=& -\sigma (x_1-y_1)\\
\nonumber
 d y_1(t) \over dt &=& -x_1 z_1-y_1+ r_1 x_1
 +\epsilon [y_2(t-\tau)-y_1(t)]\\
\nonumber
d z_1(t)\over dt &=&x_1 y_1 -\rho z_1\\
\nonumber
 d x_2(t)\over dt &=& -\sigma (x_2-y_2)\\
\nonumber
 d y_2(t) \over dt &=& -x_2 z_2-y_2+ r_2 x_2
 +\epsilon [y_1(t-\tau)-y_2(t)]\\
d z_2(t)\over dt &=&x_2 y_2 -\rho z_2.
 \label{eq:lorenz}
\end{eqnarray}

Shown in Fig. \ref{fig:direct}(b) are a few of the largest  Lyapunov exponents for  $\sigma=10, r_1=r_2=28, \rho=8/3$, showing that 
all exponents become negative abruptly at a critical delay:  hyperchaotic motion damps to a steady state. Such a transition is also observed in environmentally coupled systems \cite{resmi}, but since normal form analysis is yet to be carried out, the nature of the bifurcation remains open. 

\section{Targeting and Control} 
\label{target}

Over the last few decades the control of both periodic and chaotic oscillations in 
dynamical systems and the stabilization of unstable dynamics have been topics 
of intense research interest from both theoretical and experimental points of 
view \cite{ott,sch,tri,sinha,pyragas,pyragas1}. Several existing methods \cite{sch,tri,sinha,ott1,cont} 
stabilize fixed points by changing accessible internal parameters of a given system. 
In many natural systems, internal parameters are typically not accessible or at any rate
cannot be tuned. Thus employing amplitude death, using appropriate interactions 
between the coupled systems, suggests itself as a strategy to effect control. 

Very general methods are available for the design of specific 
fixed points in coupled nonlinear oscillators \cite{prasadnlc,prasad7}, and 
some of these can be also applied to guide dynamics onto specific limit cycles.
In this Section we will review studies that target fixed points which either exist
in the uncoupled systems or are newly created.
A related objective,  the avoidance of  \ad, the ``anti-control'' issue so to speak,
is also discussed here. 

\subsection{Steady--states through nonlinear coupling}
\label{nonlinear-target}

As discussed  in Sec. \ref{nonlinear}, nonlinear interactions can stabilize fixed points \cite{prasadnlc,prasad7}.  The strategy and application of this method to achieve new and desirable
fixed points is discussed here.  

Consider coupled oscillators with a general nondiffusive coupling function $\mathbf{G}_i$ (see Eq. (\ref{eq:model})). The essence of the procedure to target a fixed point is the following. Given a set of  desired fixed points,  ${\mathbf {\bar X}_{i}}$, these will be stationary points of the coupled system with an additional constant source, namely  of the  modified dynamical system
\beq
d \mathbf{X}_i/dt=\mathbf{F}_{i}(\mathbf{X}_i)+\epsilon \mathbf{G}_i(\mathbf{X})-
\mathbf{F}_{i}(\mathbf{\bar X}_{i}).
\eqn

The source function $\mathbf{F}_{i}(\mathbf{\bar X}_{i})$ takes a constant value
that depends on the desired fixed points. For suitable $\mathbf{G}_i$ it can  be arranged that
 $\mathbf{G}_i(\mathbf{\bar X})$ = 0. Upon variation of the  coupling parameters such as the
coupling strength $\epsilon$ (or by including time--delay $\tau$ in $\mathbf{G}_i$), 
the new fixed point can be stabilized: this, effectively, is {\sl targeted} amplitude death.

As an illustration we consider coupling between identical chaotic  R\"ossler oscillators \cite{ross} 
with exponential coupling $\mathbf{G}_i\equiv [(x_i-\beta)\exp({x_j-\delta)}, 0, 0]^T $ \cite{prasadnlc} where $T$ denotes the  transpose. The resulting equations for the coupled system are
\begin{eqnarray}
\nonumber
\dot{x}_1&=& -y_1-z_1 -\ep (x_1-\beta)\exp({x_2-\delta)}+({\bar y}_1 + {\bar z}_1)\\
\nonumber
\dot{y}_1&=&x_1+a y_1\\
\dot{z}_1&=&b+z_1(x_1-c),
\label{eq:ross1}
\end{eqnarray}\noindent
(with corresponding equations for the other subsystem). 

\begin{figure}
\scalebox{0.9}{\includegraphics{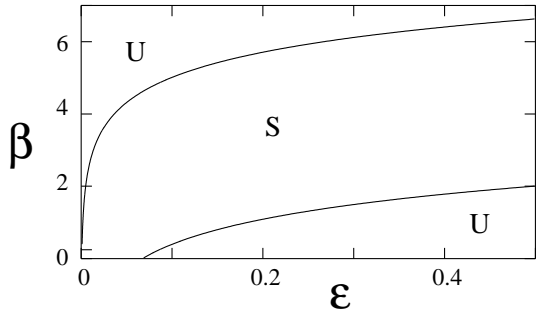}}
\caption{Schematic phase diagram in parameter space  $\ep-\beta$  for the coupled R\"ossler oscillators, 
Eq.  (\ref{eq:ross1}) \cite{prasad7}.}
\label{fig:target1}
\end{figure}

\begin{figure}
\scalebox{0.4}{\includegraphics{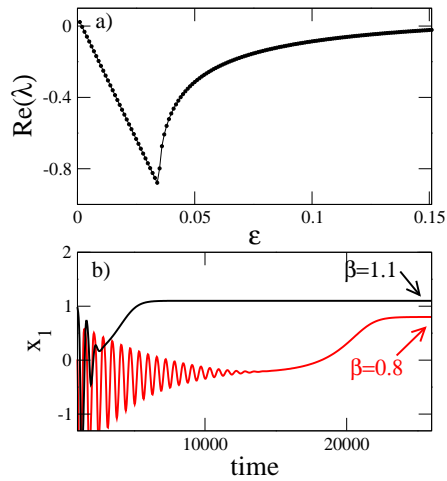}}
\caption{(Color online) (a) Real part of the largest eigenvalue (circle with solid line) as a function of the coupling strength $\ep$ 
at $\beta=1$.  (b) Transient trajectories of $x_1$  for  $\beta=0.8$ (dashed-red line) and $1.1$ (solid-black line) 
at coupling strength $\ep=0.05$ \cite{prasad7}. }
\label{fig:target2}
\end{figure}

The parameters $\beta$ and $\delta $ in $\mathbf{G}_i$  are introduced such that 
$\mathbf{G}_i = 0$ for $x_i = \beta$ and examination of the dynamical equations 
gives the fixed points ${\bar x}_{i}=\beta, {\bar y}_{i}=-\beta/a$ and ${\bar z}_{i}=-b/(\beta-c), i = 1, 2$.
The stability of this fixed point can be examined  as a function of $\ep$ and $\beta$  \cite{press}.
Stable (S) and unstable (U) regions are indicated in Fig. \ref{fig:target1}:  the unstable solution corresponds to unbounded motion, while the stable region corresponds to the possibility of the AD solution.  For $\beta=1$, the real part of the largest eigenvalue $Re(\lambda)$ \cite{press} 
at the fixed point, shown in  Fig. \ref{fig:target2}  indicates that amplitude death occurs when 
Re($\lambda$) becomes negative. Similarly the transients shown for different 
values of $\beta$ for fixed $\ep$ = 0.05 in  Fig. \ref{fig:target2}(b) arrive at different fixed points
${\bar x}_{i}=\beta$. A simple extension of this idea shows how an arbitrary point in the 
phase--space can be stabilized  \cite{prasad7}.

\subsubsection{Linear Augmentation}
\label{la}

A recent scheme \cite{manishla} (see Sec. \ref{manish}) to stabilize fixed points of nonlinear systems by coupling to a linear dynamical  system, $U$
\begin{eqnarray}
\dot{X} &=& F(X) +  \epsilon U, \nonumber \\
\dot{U} &=& -k U - \epsilon(X - B)
\label{eq:eq1}
\end{eqnarray}
\noindent
has some advantages. Since the $m$-dimensional linear system has the dynamics $\dot{U} = -k U$,
for positive $k$, and in the absence of coupling to the nonlinear system, this is incapable of having sustained oscillations \cite{manishla,resmi1}. The additional parameter $B$ in the augmented
system thus adaptively drives the $X$ dynamics to the fixed point $B$.   

This scheme is illustrated by the stabilization of arbitrary fixed points in the Lorenz oscillator \cite{lorenz}
\begin{eqnarray}
\dot{x} &=& \sigma(y - x) + \epsilon u, \nonumber\\
\dot{y} &=& r x - y - x z, \nonumber\\
\dot{z} &=& x y - \rho z,\nonumber\\
\dot{u} &=& -ku - \epsilon(x - b),
\label{eq:eq2}
\end{eqnarray} \noindent
by coupling it to a linear system as above. The fixed points when $\ep$= 0  
are ($x_0 =\pm\sqrt{\rho(r-1)}, y_0=x_0$, $z_0=r-1 $). For a the usual set of parameter values, $\sigma=10, r=28$,  
and $\rho=8/3$, these fixed points are unstable,  and the system dynamics is chaotic. 

\begin{figure}
\begin{center}
\includegraphics[width=0.4\textwidth]{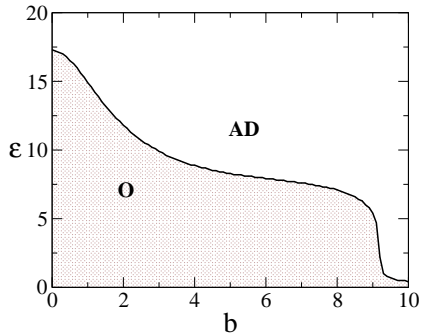}
\caption{ (Color online) Phase space diagram in parameters ($b, \epsilon$) for $k=0.01$ for $b=x_0$. 
Stable fixed point solution exists in the region marked $AD$, while $O$ represents the unstable steady state
 solution (oscillatory motion) \cite{manishla}.}
\label{fig:target-a1}
\end{center}
\end{figure}

\begin{figure}
\includegraphics[width=0.4\textwidth]{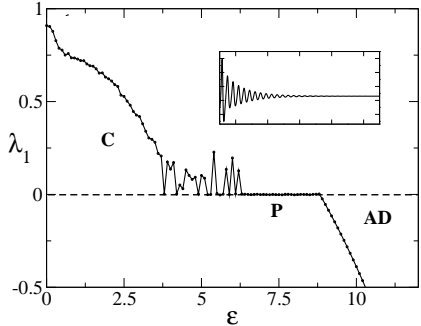}
\caption{ The largest Lyapunov exponent  as a function of coupling strength, $\epsilon$. Inset figure
  shows the transient trajectory, $x$, as a function of time  at coupling strength $\epsilon=6.8$, Eq. (\ref{eq:eq2}) \cite{manishla}.}
\vskip1cm
\label{fig:target-a2}
\end{figure}

Shown in Fig.~\ref{fig:target-a1} is a schematic phase diagram in the parameter space
$b-\epsilon$,  the shaded regime (O) representing  oscillatory and chaotic motion, but now 
with  an  (unshaded)  AD regime,  corresponding to the stabilized fixed points.
In order to achieve this we set $b= x_0 =\sqrt{\rho(r-1)}=8.4853$  and took  the decay constant $k$ = 0.01.  The largest Lyapunov exponent $\lambda_1$ is shown as a function of the coupling strength in  Fig.~\ref{fig:target-a2}(a) where the transition from  chaotic motion  to periodic motion and 
eventually to the desired fixed point  ($x_0 = y_0= \sqrt{\rho(r-1)}, z_0= r-1$). 
A typical transient trajectory of nonlinear oscillator in the AD regime at $\epsilon=6.8$ is shown in the inset. 

Similar results are also observed for other fixed points;  see Fig.~\ref{fig:target-a1} 
where AD occurs over a wider range of  $b$. In  the AD regime the coupling term 
in linear subsystem system vanishes since $\epsilon(x-b) \to  0$ as $b \to x_0$, 
reducing the dynamics to $u(t) = \exp(-kt)$ namely a decay to zero. 
This implies that after a transient,  the coupling term in oscillatory system will also 
vanish hence stabilizing the unstable fixed point $x_0$ (see the inset of 
Fig.~\ref{fig:target-a2}).  Thus {\it any} unstable fixed point can be targeted 
using an appropriate value of $b$.

By choosing $b \ne x_0$,  as the coupling terms  do not vanish the system now becomes effectively 4--dimensional (in general, the augmented system has the dimension $n + m$). In 
this case, there are new fixed points, and since the region of AD in Fig.~\ref{fig:target-a1} is 
quite wide, these can also be stabilized. Of course, this is distinct from targeting; see details in \cite{manishla}.

\subsection{Reprieve: Avoiding fixed points}

Some situations demand that the fixed point solution be unstable, and either  chaotic or periodic solutions should be targeted.  Although it is easy to perturb the system by external stimuli to avoid \ad, getting a specific solution is not always simple. Here we discuss methods that specifically attempt to find particular periodic solutions through targeted nonlinear interactions; see Sec. \ref{nonlinear-target}.

\begin{figure}
\scalebox{0.7}{\includegraphics{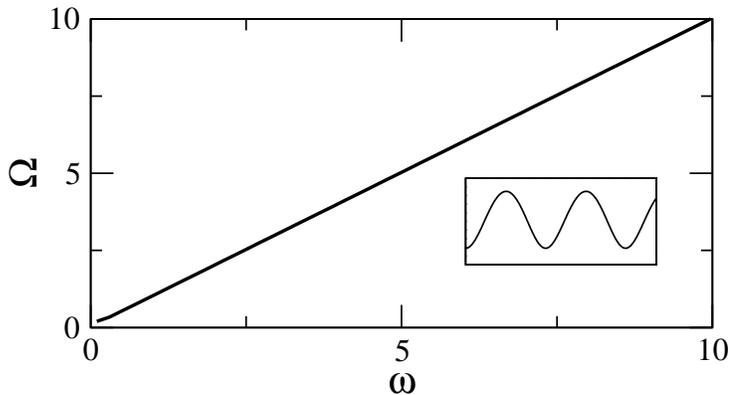}}
\caption{Variation of the common frequency of coupled R\"ossler oscillators ($\beta=0.4$ and $\epsilon=0.01$)
as a function of   targeted frequency $\omega$ \cite{prasad7}.}
\label{fig:target-p}
\vskip.1cm
\end{figure}

A modification of the coupling function $G$, for instance making it $ \{x_i-\beta  \sin(\omega t)\}$ in Eq. (\ref{eq:ross1}) \cite{prasad7} results in the dynamics being on a periodic orbit of 
frequency $\omega$. This behaviour is robust, namely it occurs in a specific range of parameters. If one plots the frequency of the synchronized R\"ossler oscillators  as a function of the forcing as in
Fig. \ref{fig:target-p}, one can see that the  common  frequency $\Omega$ 
can be made to assume {\it any} desired value. Note that the other parameters are fixed so that in the absence of periodic forcing, amplitude death would result. The input and output frequencies are 
identical, showing that one can indeed target periodic motion as desired. 
The inset figures in Figs. \ref{fig:target-p} show  the   time series ($x_1$ vs time) 
associated with such a targeted periodic motion for $\omega=5$.

As indicated earlier, this method ensures only that the frequency of the targeted periodic solution takes a specific value;  controlling the amplitude of the oscillations remains an open problem \cite{prasad7}.

\subsubsection{Gradient coupling} 

The gradient (or directional)  coupling introduced recently \cite{zougc} for $N > $2 is a new and interesting way to avoid AD. Note that amplitude death in time-delay coupled systems  occurs in ``islands'' in parameter space (the parameters in question being the delay time $\tau$ and the coupling strength $\epsilon$; see  Sec. \ref{delay}). In order to avoid such islands of AD asymmetry in the coupling strength, keeping the  coupling function the same has been used. Consider the modification in  Eq. (\ref{eq:reddy}) as 
\begin{eqnarray}
\dot{Z}_{1}&=&[1+i\w_{1}-|Z_{1}|^2]Z_{1}+(\epsilon+r)[ Z_{2}(t-\tau) -Z_{1}],\nonumber\\
\dot{Z}_{2}&=&[1+i\w_{2}-|Z_{2}|^2]Z_{2}+(\epsilon-r)[ Z_{1}(t-\tau) -Z_{2}].
\label{eq:gc}
\end{eqnarray} \noindent
The asymmetry parameter $r$ is introduced in the coupling so that  when $r$ = 0 the system is homogenous,  while when $r = \epsilon$ the coupling is anisotropic, leading to one--way coupling \cite{zougc}. 
Such coupling arises in many practical situations \cite{zougc}.

\begin{figure}
\scalebox{1.5}{\includegraphics{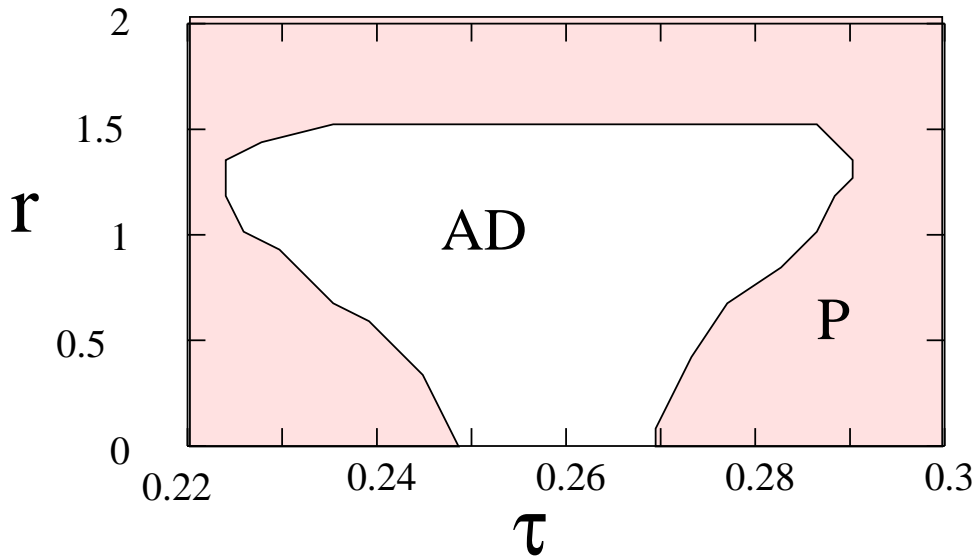}}
\caption{The schematic phase diagram  in parameter space $\tau-r$ for fixed coupling strength $\epsilon=2$, Eq. (\ref{eq:gc}).}
\label{fig:gradient}
\vskip.1cm
\end{figure}

Shown in Fig. \ref{fig:gradient} is a schematic phase diagram in $\tau-r$ space for fixed coupling strength $\epsilon=2$. For $r$ = 0, the system behaves as shown in Fig. \ref{fig:reddy1} where AD is observed within the interval $\tau \in$ [0.248, 0.27].  When $r$ is increased,  the islands of AD get truncated (although the behaviour is nonmonotonic)  showing that AD can be avoided in
through interactions that have directionality. A detail analysis can be found in \cite{zougc}.

Konishi \cite{konishi9} has shown that when the Jacobian matrix at a fixed point of an isolated system has an odd number  of real positive eigenvalues,  then time-delay connections are not able to induce \ad~ to this fixed point. It is clear that such a property can be designed, possibly through  augmentation of the system  (Sec. \ref{manish} \& \ref{la}) into  higher dimensions. 
This latter approach and the  gradient method discussed above are of use when there is  time--delay coupling.  It should be emphasized that there is no general method that can be
used in order to avoid  \ad~ from any scenario, and thus a techniques for avoiding AD are
necessary, and the problem is worthy of further study.

\section{Characterization}
\label{character}

Since the asymptotic dynamics in the region of  \ad~ is always stationarity, the interesting dynamics
are  essentially the transient behaviour.  Since this is also always a result of interaction between
two or more systems, other phenomena that arise in coupled dynamics such as  synchronization, riddling, and multistability can (and often do) occur, and should be taken into consideration in  understanding the nature of dynamics near fixed points. 
In this section we focus on  the nature of transient dynamics through analysis of the 
Jacobian matrix at the fixed point. AD can only occur when all eigenvalues have negative
 real parts, but the nature of the transients depend on whether there is a complex component
 or not.  
 
\begin{figure}
\scalebox{.5}{\includegraphics{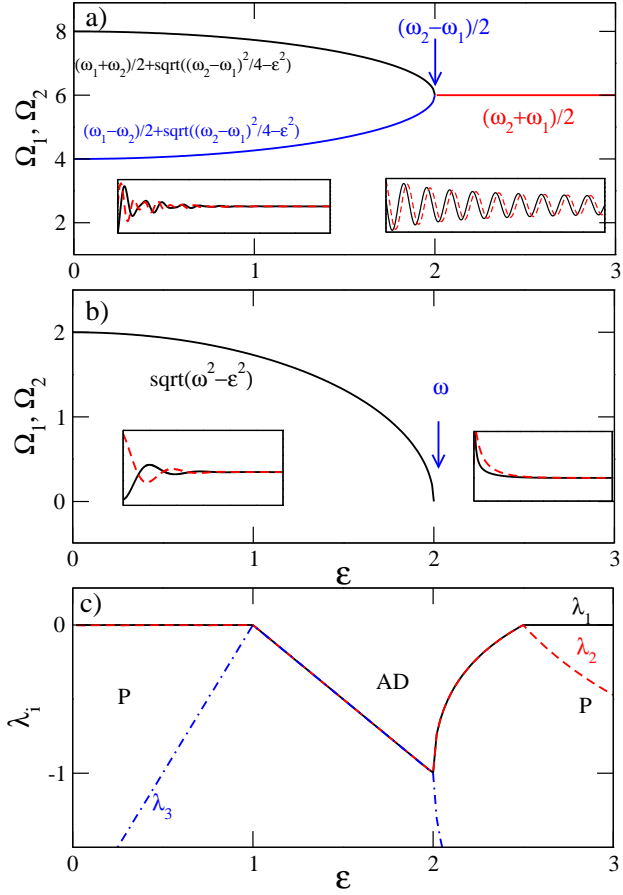} }
\caption{(Color online) Frequencies, $\Omega_1$ and $\Omega_2$, for respective oscillators for (a) mismatched, $\omega_1=4$ and $\omega_2=8$, 
Eq. (\ref{eq:ls}) and (b) counter rotation, $\omega_1=2$ and $\omega_2=-2$ (Sec. (\ref{mismatch})).
(c) Largest three Lyapunov exponents for counter rotation.
Inset figures show the transient trajectories in respective regimes as a function of time at coupling strengths $\ep=1.5$ and $\ep=2.2$.}
\label{fig:freq}
\end{figure}

We first consider mismatched Landau--Stuart oscillators (Sec. \ref{mismatch}) without time--delay.
Shown in Fig. \ref{fig:freq}(a) is the frequency of individual oscillators as a function of the 
coupling strength keeping other parameters as in  Fig. \ref{fig:reddy1}. Straightforward analysis
of the characteristic equation for the eigenvalues after setting $\lambda = \alpha+ i \beta$
and separating real and imaginary parts gives, for the leading eigenvalue, the  result
\begin{eqnarray}
\alpha&=&1-\ep  \\
\label{eq:ev1}
\beta&=&(\omega_1+\omega_2)/2\pm\sqrt{(\omega_1-\omega_2)^2/4-\ep^2}
\end{eqnarray}
\noindent
if  $\ep <(\omega_1-\omega_2)/2$, while
\begin{eqnarray}
\alpha&=& 1-\ep +\sqrt{\ep^2-(\omega_1-\omega_2)^2/4}\\
\beta&=&(\omega_1+\omega_2)/2
\label{eq:ev2}
\end{eqnarray}
when $\ep >(\omega_1-\omega_2)/2$.

This result holds when the direction of ``rotation'' of both oscillators is the same, which here 
corresponds to $\omega_1$ and $\omega_2$ having the same sign \ie
if one oscillator (in Fig. \ref{fig:model}(a)) are rotating in clockwise then another follows  the same direction. When the signs differ, the analysis is more involved, and for the case $\omega_1= - \omega_2= \omega$ (see details in Ref. \cite{prasad1}) the real part of
 the eigenvalue is unchanged  (Eqs. (\ref{eq:ev1}) and (\ref{eq:ev2})) but 
 \begin{eqnarray}
\beta&=&\sqrt{(\omega^2-\ep^2)}
\label{eq:ev1c}
\end{eqnarray}
\noindent
for $\ep <\omega$, while $\beta$ = 0 if  $\ep >\omega.$

Shown in the insets of  Figs. \ref{fig:freq}(a) and (b) are the transients  for both cases, namely 
co- and counter-rotating coupled oscillators.  Fig. \ref{fig:freq}(c) shows the occurrence of AD 
when the signs of $\omega_i$ differ; compare the behaviour seen in Fig. \ref{fig:aronson1} 
for the same extent of parameter mismatch.  In Fig. \ref{fig:freq}(a) note that there is no synchrony
for  $\epsilon=(\omega_1-\omega_2)/2$,  but for counter-rotating oscillators, the motion is
synchronized for $\epsilon < \omega$ while beyond this value of the coupling the system is
{\sl overdamped}.  
These results suggest that the nature of the transients thus  depends on the scenarios 
(Sec. \ref{scn}) as well as on whether the transition is from a globally synchronized
state or not. 

\subsection{The phase--flip}

\begin{figure}
\scalebox{.5}{\includegraphics{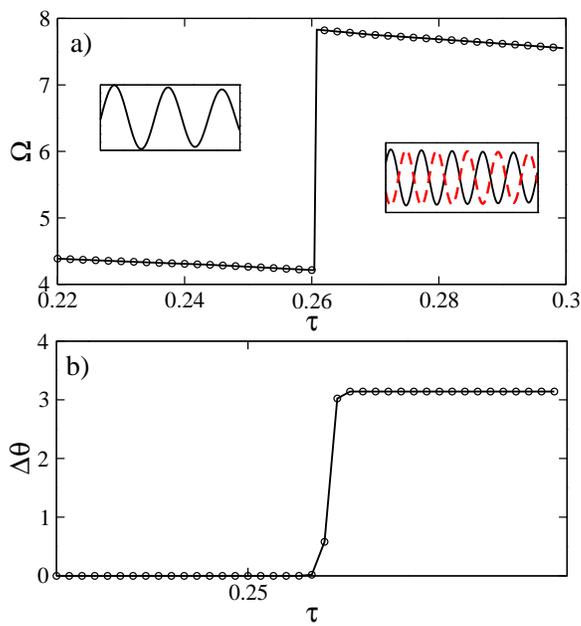} }
\caption{ (Color online) (a) The common frequency of oscillations, (b) phase difference between oscillators for time-delay coupling, 
Eq. (\ref{eq:ls}).
The inset figures show the transient trajectories in respective regimes as a function of time at time-delays $\tau=0.24$ and $\tau=0.28$.}
\label{fig:freq-delay}
\end{figure}

When identical oscillators are coupled with time--delay  (Sec. \ref{delay}) 
they first synchronize in phase. Shown in Fig. \ref{fig:freq-delay}(a)
is the common frequency and in (b) the phase difference between the oscillators.
Within the AD region, there is a point at which there is an abrupt change in the 
frequency and this is accompanied by a jump in phase difference by $\pi$: the transient 
trajectories (inset figures) go from being synchronized in-phase to being synchronized
out-of-phase behavior.

This transition, termed the phase--flip has been studied in detail in Refs.
 \cite{prasad3,prasad4,appfb2,rajat2,rajat3} in a variety of systems,  as well as in 
different dynamical regimes.
Furthermore, when there are more than two coupled oscillators,
the phase--difference is not necessarily $\pi$, but can depends upon the specifics of the system 
as well as  on the coupling topology  of the network \cite{ap3}. For asymmetric time--delays, even 
in the case of two coupled oscillators, the phase difference can be different from $\pi$
\cite{rajat3}. Such different phase relations have been seen in experiments with 
asymmetric coupling \cite{yoshimoto1} and time-delay \cite{rajat3} as well as in
in relay--coupled nonlinear oscillators \cite{manish3}.

As we have seen in Sec. \ref{multi} where  \ad~ coexists  with periodic motion, multistability 
can occur prior to the onset. The possibility of coexistence of both chaotic dynamics as well
as stationarity has been observed in mixed R\"ossler and Lorenz systems \cite{prasad3} where there is also riddling. Although multistability is not a property of fixed point here, it should also be
emphasized that at a given set of parameters reaching the fixed point solution can 
depend on initial conditions as well.

\section{Amplitude death on networks}
\label{network}

Most of the systems discussed in the previous sections are of two coupled systems only. However, many
systems of interest are  composed of large numbers of interacting oscillators---particularly in biological contexts---which are coupled on networks.  Mathematically a network is a graph where $N$ nodes (or vertices) are connected in a specific manner. The details of the connections are an input in $\mathbf{G}$ (see Eq.  (\ref{eq:model})).  There is considerable interest in networks 
in diverse areas of the sciences, spurred by the realization that these ideas have wide applicability  
 \cite{albert}. The internet, food webs, neural networks,
electrical power grids, coauthorship and citation networks of scholars, biological networks are well known instances \cite{albert,network1,network2}. 

The nature of the complex dynamical behavior on networks has a sensitive dependence on the 
nature of the connections between the nodes in addition to the actual dynamics of the 
individual subsystems.  The occurrence of  AD in networks of coupled oscillators  \cite{rub,atay4,yang,erm1,wangn,chenn,wang}  has been investigated for a variety of topologies ranging 
from small--world connections \cite{hou} to scale--free  networks \cite{liu} and the 
ring topology \cite{sen,konishi10}. AD as an emergent phenomenon has been seen in 
networks of  coupled maps \cite{atay2} and in chaotic coupled map lattices \cite{konishi4}. 

Since one important area for the occurrence of AD is in neuronal networks, we study here 
a system composed of identical Hindmarsh--Rose (HR) neurons \cite{hr} that are 
synaptically  
coupled in different topologies,
\begin{eqnarray}
\nonumber
\dot{x}_{i}&=&a x_i^2-x_i^3-y_i-z_i-\frac{\epsilon}{K_i} \sum_{j = 1}^{K_i} A_{ij} h(x_i,x_j,\tau)\\
\nonumber
\dot{y}_{i}&=&(a+b)x_i^2-y_i\\
\dot{z}_{i}&=&c(dx_i+e-z_i)
\label{eq:hr}
\end{eqnarray}
where
\begin{eqnarray}
h(x_i,x_j,\tau) =  \frac{(x_i-V_s)}{[1+\exp \{-\beta (x_j(t-\tau)-\Theta_s)\}]}.
\label{eq:hrG}
\end{eqnarray}\noindent
The topology is specified through the adjacency matrix elements, $A_{ij}$, which takes
the value 1 if the nodes are connected, and 0 if they are not. 
At each node $i$ one has a neuron, and the notation is as follows:  
 $x_i$ is the membrane potential  and $y_i$ and $z_i$ are the  
fast and slow currents respectively. The external parameters are the 
synaptic coupling  strength $\ep$ and delay $\tau$. Here the reversal 
potential  $V_s$ is fixed at $V_s=2$ at which the synapse is excitatory, 
the spiking threshold is fixed at  $\Theta_s=-0.25$ and the  synaptic 
coupling function is taken to be sigmoidal. The other parameters are 
set to standard values,  $a = 2.8, b =1.6, c = 0.001, d = 9$ and  $e = 5$ \cite{hr1}.  

\begin{figure}
\scalebox{0.35}{\includegraphics{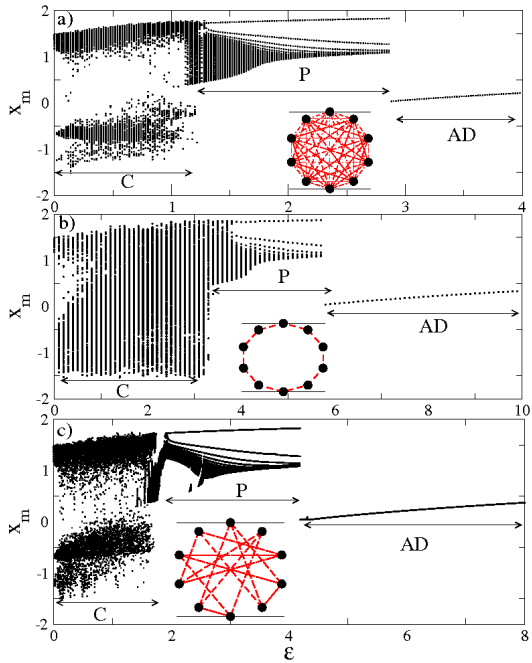}}
\caption{(Color online) Bifurcation diagrams  for the system of $N$= 10 coupled 
oscillators with (a) global coupling, (b) nearest neighbor coupling with periodic boundary conditions,
and (c) a small--world network where each neuron is randomly connected to two others (shown in the insets).
$x_m$ corresponds to the maxima of $x_1$ in Eq. (\ref{eq:hr}) (see for details Ref. \cite{prasadnlc}).}
\label{fig:network}
\end{figure}

Networks of such oscillators coupled without time--delay (namely for $\tau=0$) 
continue to show  global amplitude death as does the case of two oscillators. 
 Fig \ref{fig:network}  is  bifurcation diagrams for networks of $N$ = 10 oscillators 
 coupled in different topologies. Three cases  considered for illustration include 
 (a) global connection, namely  $K_i = N$-1 in Eq. (\ref{eq:hr}) and all $A_{ij}$= 1
 for $i \ne j$, 
(b) a linear chain with periodic boundary 
conditions,  $K_i$ = 2 as well as (c) random connections with $K_i$ = 3.
The motion goes from being oscillatory to amplitude death as  the coupling strength
 is increased in each of the cases, although the threshold depends on the 
 coupling topology. In general, though, the approach to 
the amplitude death regime is gradual with the range of oscillation 
becoming narrower with increasing
coupling. 

The behavior of large ensembles of coupled weakly nonlinear oscillators has been 
studied by Ermentrout and coworkers \cite{erm1} in the finite large $N$ as well as the $N \to \infty$ limits. The state of uniform rest can become stable when the coupling is sufficiently strong 
and the frequencies are sufficiently disparate. Amplitude death is also reported in limit cycle oscillators coupled via mean field coupling and randomly distributed 
frequencies \cite{strogatz1,strogatz2,strogatz3,erm2}. When the distribution of frequencies has sufficiently large variance, the oscillators effectively pull each other off their limit cycles 
into the origin which is a stable equilibrium point for the coupled systems.

A closed chain of delay coupled identical limit cycle oscillators that are near 
supercritical Hopf bifurcation \cite{sen} also show AD. The model is a 
discrete dynamical system, and a dispersion relation valid for an arbitrary number of oscillators
is obtained in the continuum limit. A detailed linear stability investigation of these states has
been carried out in order to delineate their actual existence
regions and to determine their parametric dependence on time delay. 
It is shown that island regions of AD can exist for any number of oscillators so long as the delay is  finite: in effect, time delay contributes toward their stability. 
The size of these island is independent of $N$ when $N$ is even and is a 
decreasing function of $N$ when $N$ is odd. Sen and coworkers  \cite{sen,reddy2,reddy4}
have extensively investigated \ad~ in delay--coupled networks. 

By applying averaging methods to diffusive time--delay coupled 
networks of weakly nonlinear oscillators \cite{atay4}, it is possible to 
determine parameter ranges in which the network experiences 
amplitude death \cite{cheng2}. Time--delay also suppresses oscillations 
in network of oscillators coupled in a one--way ring topology \cite{konishi10}.
When two networks are coupled \cite{cheng2} delayed effects of shortcuts
in each network and coupling between the two groups are considered. 
When the short-cut delay is removed the death of oscillations is caused by 
variational coupling delay. This behavior is found to hold in  large number of
locally coupled oscillator as well.

As already indicated, AD in neuronal networks is a phenomenon of practical interest,
and there is some indication that such dynamics is significant in understanding 
the temporal activity appearing in the olfactory bulb \cite{monteiro}.
In this context there has been an analytic study of neuronal models on a network with time--delayed self connection. The effect of delay on the stability  of the  trivial solution, 
and on the existence of self--sustained periodic solutions has been investigated \cite{monteiro}. 
Neural networks with distributed delay also show \ad~\cite{omi}. When the delay distribution
has dispersion below that of the  exponential distribution, the system exhibits reentrance:
stability is lost and then recovered when the mean delay is increased. If the delay dispersion
is greater than that of the exponential distribution, the system never destabilizes.

Partial \ad---effectively a chimera---has been seen in a small network of globally coupled oscillators \cite{ling}. 
Konishi and coworkers \cite{konishidc1,konishi10,konishidc2,konishidc3} have studied
a number of models with different types of connections, on networks of varying topologies. 
Analytical estimation of regimes of  \ad~ have been carried out, and their analysis 
shows that the odd--number property that is known in delayed feedback control
also exits in global dynamically coupled oscillators.  Furthermore global amplitude death is experimentally demonstrated in a Chua's circuit coupled by an RC line.  Various transitions 
including amplitude death have been also investigated in a regular array of 
nonlinear oscillators \cite{yang}.

A systematic procedure for achieving \ad~involves adjusting the coupling strength 
and delay times in the connections \cite{konishi4}. Such a procedure has the advantage
that stabilization can be achieved independent of the network topology. This has been
applied with success to the well--known double-scroll circuit 
on a small-world network (a realization can be seen in Fig. \ref{fig:network}(c)). 
Hou \etl~ \cite{hou} investigated oscillator death behavior on small--world 
networks and found that small-world connectivity can eliminate the 
AD seen for the same model on a  regular lattice. On the other hand, the
small-world connectivity can also lead to global oscillator death which is absent 
in the regular lattice or for that matter on a completely random (Erd\H{o}s--Renyi) network.  
On scale-free networks of nonlinear oscillators AD has been investigated both 
numerically and analytically in \cite{liu}. 

Networks of coupled scalar maps also show \ad~ as shown by Atay \cite{atay3} 
who studied such a network with weighted connections taken so that they may 
include time--delay. The emergence of AD in a chaotic coupled map lattice 
with  irregular network topologies \cite{konishi11}
or on networks with globally dynamical interaction \cite{konishidc2} has been 
investigated in recent work. Similarly, amplitude death that 
results from conjugate coupling has been verified theoretically in a network of chaotic 
oscillators under local and global conditions \cite{zhang}. 

The  Belousov-Zhabotinsky reaction in a continuous-flow
stirred tank reactor (CSTR) is a well--defined nonlinear oscillator that can be realized in experiment. 
In addition to the study of theoretical models, experiments on networks of real systems like these also report \ad~\cite{yoshimoto2}. In the next Section we discuss several experimental systems where
AD can be realized.

\section{Experiments and applications}
\label{experiment}

The stabilization of steady states, either those corresponding to the uncoupled
oscillators or those created by design has been observed in various experiments.
This has a variety of potential applications, and we discuss these here. 
  
Early experimental demonstrations of AD were in  coupled Belousov--Zhabotinskii (BZ)
reactions \cite{eli5}, namely chemical oscillators.  Other experiments on 
asymmetrically coupled BZ reactions \cite{yoshimoto1} also lead to AD.
The experimental set up used identical BZ reactors  coupled via mass  transfer- 
the coupling is thus  diffusive in nature and since it depends on the 
concentration gradient both symmetric and asymmetric coupling can be realized 
by changing the active volumes of the tank reactors.  This  is of importance in 
biological systems where systems are indeed coupled through the exchange of fluid
and are likely to be asymmetric in nature.  A situation that is particularly amenable to experimental
control is that of electrochemical oscillators \cite{zhai}; by coupling nearly identical
systems in the proximity of  their individual Hopf bifurcation points, they can be 
driven to AD \cite{aronson}  as discussed in  Sec. \ref{mismatch}.

Thermo--optical oscillators \cite{herrero},
where the coupling is linear and by heat transfer are another class of coupled 
systems that show AD.  Thermo kinetic oscillators \cite{zeyer} that show temperature and concentration oscillations are of direct relevance in industrial processes.  
Most  of these are exothermic in nature, and AD, which has been seen experimentally 
in such systems can have severe consequences; these studies thus may have a
significant impact on safety measures \cite{zeyer}. 

Many systems and scenarios for AD have been experimentally investigated in 
coupled electronic circuits. Apart from the Landau--Stuart system, the van
der Pol oscillator and other standard nonlinear or chaotic oscillators that are 
easily realized in analog circuits, a number of hybrid models can also be studied
in experiment. This extends the study of mismatched oscillators beyond the cases
of dynamic coupling  \cite{konishi10} or 
when the frequencies of coupled units differ \cite{setou,okamoto} to the situation 
when the coupled systems are completely distinct from each other: the interaction 
between an  electronic and biological oscillator---a neural--electronic hybrid interface---can also suppress oscillatory dynamics \cite{ozden} .  

Partial \ad~is the situation when instead of all variables reaching a stationary 
state, only some of them do \cite{partial}. This behaviour is seen when strong 
interactions break down both reflection and  translational symmetry in the coupled 
systems, and has been experimentally verified using  coupled 
Lorenz circuits \cite{qing}. Partial AD is distinct from the case of the chimeric behaviour in 
networks, where some members of a set of coupled oscillators remain oscillating while
others go to a stationary state \cite{yang,atay2,konishi13,volkov}.

Experiments on coupled lasers have been a major area for the application of many ideas
discussed above since delay coupling can be easily implemented.  Pyragas \etl~ 
\cite{pyragas2} have proposed that the stabilization of fluctuations in multi-mode 
intracavity doubled diode pumped Nd:YAG lasers can be achieved by stabilizing the 
steady state, namely through AD. A related experimental study of chaos control 
in lasers by Bielawski \etl~ \cite{bielawski}  stabilizes the unstable steady state of
a fiber two-level class B laser.  Delayed coupling with delayed  feedback 
has been shown to induce AD in a laser system \cite{vicente}, and   Roy and 
coworkers \cite{ kim1,kim2} have experimentally verified a number of issues in the 
dynamics of coupled systems, including \ad. 

The fact that several meteorological phenomena are oscillatory and are frequently coupled suggests that AD will also occur in climate models. There are large--scale oceanic and atmospheric decadal anomalies. This  variability is characterized by patterns of sea--surface temperature anomalies, correlated with the atmospheric  pressure at sea level \cite{gallego}. In the ocean, this
variability results in fluctuations of ocean temperature and ocean mass transport. In the atmosphere 
these fluctuations are associated with changes in 
surface westerlies and surface air temperature over ocean basins and the  
period of these oscillations is appears to depend on the transit time of oceanic Rossby waves;
these are of the order of decades. Gallego and  Cessi  \cite{gallego}, using their model for mid-latitude large scale interaction between the upper ocean and the troposphere consider the case when two ocean basins are coupled through zonally average atmosphere.
Each ocean basin also interacts with the atmosphere via wind driven torques and heat fluxes at 
the sea surface. When uncoupled,  each individual ocean basin has sustained oscillations, but with coupling, the oscillatory anomalies decay in time, and the system eventually reaches a steady state.

This vital result in the study of climate change is thus linked to \ad~which occurs only when the long Rossby wave delays for the two ocean basins are sufficiently different. The  model formulated here uses  the  Rossby wave delay times to control the natural frequency of oscillation in each basin, 
and thus the case of AD seen here can be seen as an example of mismatched 
interacting oscillators as discussed in Sec. 2.1.

Other large--scale systems that show AD include  ecological models as for example coupled 
prey predator systems \cite{eurich}. Amplitude death  is important in epidemical models as well, \eg in infectious disease, its important to predict whether the infection  has disappeared or the pathogen persists. Steady states in the models of pathogen--immune dynamics of infectious diseases  \cite{neamtu} are other instances of systems where the AD state is a desirable end point, and thus methods of control, discussed in Section 4 are important in the context of intervention strategies. 

The issue of \ad~has been also discussed in the field of neural oscillators (See Eq. (\ref{eq:hr}) and corresponding Fig. \ref{fig:network}).   Amplitude  death in a neural network model \cite{he}
where the interaction is ``transferable'', namely  in a cyclic chain of oscillators with one--way coupling  and with time lag.  Amplitude death  has also been observed in a network of 
Fitzhugh--Nagumo neurons \cite{gassel}.  Diffusive coupling is not the only means of achieving \ad. Phase--repulsive coupling \cite{ullner}  also eliminate oscillations in a population of synthetic genetic clocks.

\section{Summary}
\label{summary}

The phenomenon of amplitude death  has attracted considerable attention in the past decades
following the realization that this is quite general and that this has significant practical implications. 
 
In this review we have shown that AD is a general outcome in coupled nonlinear oscillator systems,
occurring in a wide variety of dynamical models and for different forms of coupling.  There are a number of scenarios which lead to amplitude death and we have tried to use the well--studied Landau--Stuart limit cycle oscillator model to illustrate these. A detailed theory---in terms of bifurcation structure---is not yet available, and this constitutes an interesting and  open problem.

The asymptotics of AD is a featureless steady state, but the transient dynamics can be of interest, with the motion of the coupled systems having nontrivial phase relations.  The frequency of the damped oscillations can also show a discontinuity: this is the phase--flip  transition. 
We have reviewed here these various dynamical characteristics in model systems. 

Targeting a specific steady state is a central issue in the study of nonlinear dynamics, particularly with regard to stabilizing  low--dimensional dynamics such as fixed points or periodic orbits. We have reviewed some aspects of targeting fixed point solutions by considering specific interactions. 
This will be essential if one is to be able to use the phenomenon of AD in practical applications. 
Targeted fixed power outputs  in coupled laser systems can have significant applications  in laser technology \eg laser surgery, or laser welding and fabrication. 
In other practical applications AD is to be avoided, for instance in brain function; we have also reviewed methods for the reliable avoidance of AD. There can be significant constraints, though since many natural systems have specific and constrained coupling forms---for instance the synaptic 
coupling between neurons, or the diffusive interaction between nonlinear chemical systems.  

Although we have considered  simple systems for purposes of  illustration, AD is known to occur in higher dimensional systems as well  \cite{stich,saka,neu}. Time--delay makes the dynamics effectively  infinite--dimensional  and thus there can be AD in hyperchaotic systems.

External noise is unavoidable in natural systems or in experiments. AD is quite robust with regard to noise since the Lyapunov exponents of systems undergoing AD are substantially negative; the  systems are deep in the region of stability.  Thus AD has  been observed in numerous experiments  many of which have been  discussed here.  This aspect deserved further study, however, since it is not clear to what extent inhomogeneous steady states can be perturbed by noise (\eg after $\epsilon=2.5$ in Fig. \ref{fig:multi}). In addition, the study of  network is currently of great importance in various areas of study and as we have seen, AD in complex  systems with different topologies is of interest. The chimeric state of partial amplitude death is, in particular, an important one that needs further exploration.

Although there has been substantial work on amplitude death,  several challenges
remain both at fundamental as well as applications levels.  Through this review we have tried  to
summarize the  current state of understanding of the phenomenon, as well as  drawn attention
to several open issues, and to highlight possible applications.

\section{Acknowledgments}
GS is supported by the Council of Scientific  and Industrial Research,
India through a Senior Research Fellowship. AP acknowledges the support
of Delhi University and the kind hospitality and financial support from the
MPI-PKS Dresden, Germany. We are grateful to R. Karnatak, N. Punetha,
M. D. Shrimali, S. K. Dana, and J. Kurths for useful discussions and
collaboration over many years, and the Department of Science and Technology,
Government of India, for research support.

\end{document}